Stephen J. Pennycook, Changjian Li, Mengsha Li, Chunhua Tang, Eiji Okunishi, Maria Varela, Young-Min Kim & Jae Hyuck Jang

# Materials Structure, Properties and Dynamics through Scanning Transmission Electron Microscopy

## Abstract

Scanning transmission electron microscopy (STEM) has advanced rapidly in the last decade thanks to the ability to correct the major aberrations of the probe forming lens. Now atomic-sized beams are routine, even at accelerating voltages as low as 40 kV, allowing knock-on damage to be minimized in beam sensitive materials. The aberration-corrected probes can contain sufficient current for high quality, simultaneous, imaging and analysis in multiple modes. Atomic positions can be mapped with picometer precision, revealing ferroelectric domain structures, composition can be mapped by energy-dispersive X-ray spectroscopy (EDX) and electron energy loss spectroscopy (EELS) and charge transfer can be tracked unit cell by unit cell using the EELS fine structure. Furthermore, dynamics of point defects can be investigated through rapid acquisition of multiple image scans. Today STEM has become an indispensable tool for analytical science at the atomic level, providing a whole new level of insights into the complex interplays that control materials properties.

1. Introduction

With the successful correction of lens aberrations, the STEM has become the dominant microscopy technique used today in materials research, due to the availability of simultaneous, multiple, imaging and spectroscopic modes. While these benefits have long been appreciated in principle (Crewe 1966; Crewe et al. 1970; Rose 1974), before aberration correction it was difficult to get sufficient current into the probe for good quality images, nor could spectroscopic signals be obtained at atomic resolution. Aberration correction, bringing smaller, brighter probes, has overcome the historic disadvantage of STEM, that of poor signal to noise ratio.(Pennycook and Nellist 2011) In this review we highlight some recent achievements and applications to materials. More detailed accounts can be found in a number of recent reviews. (Pennycook 2015; Oxley et al. 2016; Varela et al. 2017; Gazquez et al. 2017; Li et al. 2017)

Figure 1 shows the principle of STEM. Like a scanning electron microscope, an incident probe is scanned across a sample, but it is thin enough so that the beam is transmitted, then several signals can be detected simultaneously and used to form images with complementary characteristics. The high angle annular dark field (HAADF) detector collects Rutherford scattering from the atomic nuclei, producing an image with strong sensitivity to atomic number Z, often called a Z-contrast image. Light columns such as O are only weakly visible in the Z-contrast image, but are seen clearly in a simultaneous bright field (BF) image or annular bright field (ABF) image. In perovskites and related materials this allows the octahedral rotations to be determined, which are crucial to understanding their properties. If the bright field detector is removed, electrons can be passed through a magnetic sector electron energy loss spectrometer to provide elemental maps and electronic structure information.

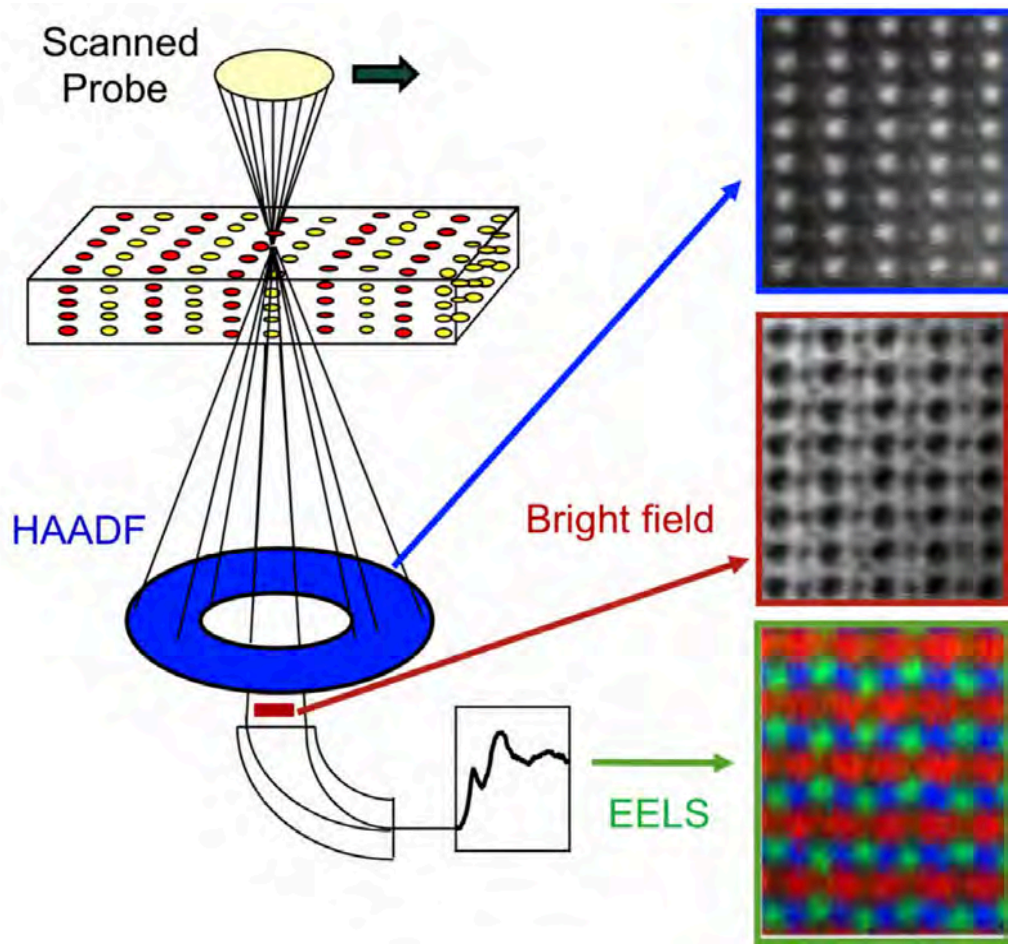

Fig. 1. Schematic showing the multiple imaging and analytical signals available in STEM. Top right is a HAADF image of $BiFeO_3$; the brightest atom columns are Bi and the less bright ones Fe, which are seen displaced slightly to the right due to the ferroelectric polarization. The centre image is a bright field image showing O columns, below is a color composite EELS image of $LaMnO_3$ in which the Mn is shown in red, O in green, showing displacements due to octahedral rotations, and La in blue. Adapted with permission from Borisevich et al. 2010b, copyrighted by the American Physical Society, and Varela et al. 2012.

2. Imaging and Spectroscopic modes

2.1 Imaging Modes

The HAADF image is an incoherent image, that is, a direct image, of the columnar scattering power.(Pennycook and Boatner 1988; Pennycook and Jesson 1990; Pennycook and Jesson 1991) It is least sensitive to crystal tilts, specimen thickness and residual aberrations, therefore it can provide positions of high-Z columns with high accuracy, as demonstrated in Fig. 2 showing $BiFeO_3$ (BFO) viewed along the pseudocubic [110] direction (denoted $[110]_{pc}$). However, for O columns, which scatter weakly to the HAADF detector, a bright field image gives more accurate positions, although, being a phase contrast image, the correct defocus and thickness combination must be determined by image simulation. Combining the positions from both (simultaneously acquired) images allows the octahedral rotations to be accurately measured. (Wang et al. 2016a; Kim et al. 2017)

The annular bright field image, a phase contrast image first proposed by Rose, (Rose 1974) has also become popular in the aberration corrected era for

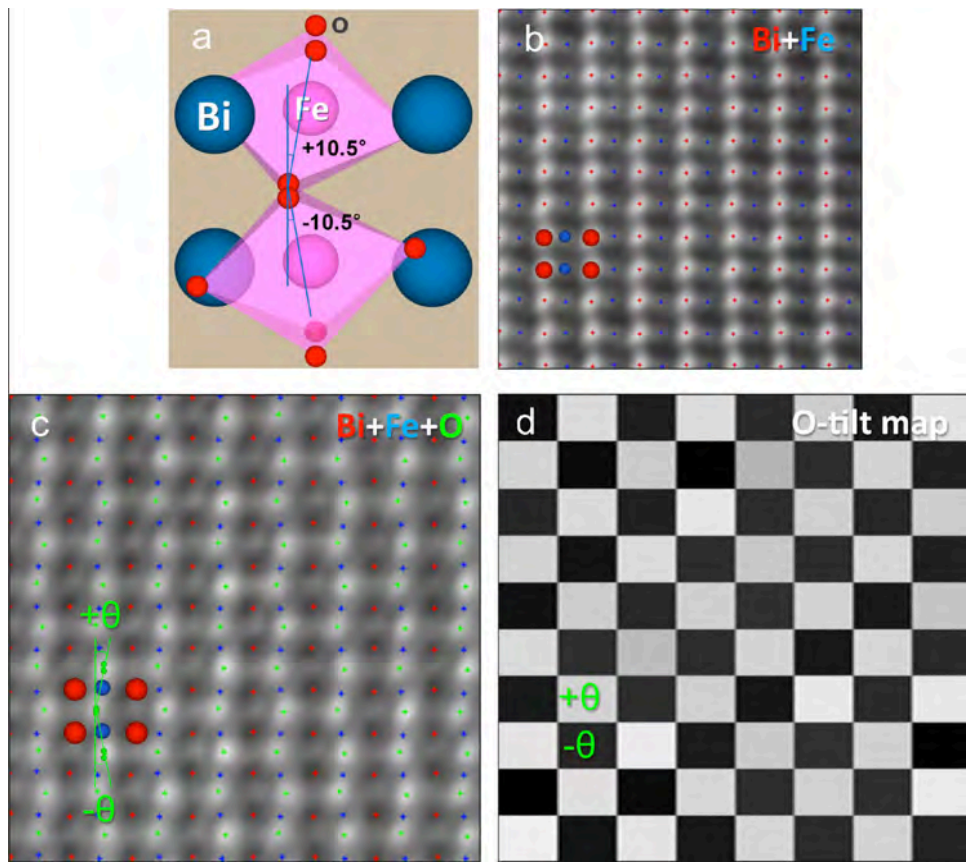

Fig. 2. Measurement of oxygen octahedral tilts in BFO using HAADF and BF STEM images. (a) Schematic of the BFO structure observed along the $[110]_{pc}$ direction. (b, c) Simultaneously acquired HAADF and BF STEM images overlaid with peak finding results for the atomic columns. Bi (red) and Fe (blue) atomic columns can be defined from the HAADF STEM image, (b) and oxygen columns from the BF STEM image, (c). (d) A checkerboard pattern generated from the measurements of the projected tilt angles ($\pm \theta$) of oxygen octahedra. Reproduced from Kim et al. 2017. Copyright 2017 with permission from Elsevier.

imaging light atoms.(Findlay et al. 2010; Ishikawa et al. 2011) However, the annular bright field image may be more sensitive to crystal tilts than a normal bright field image, as seen from the comparison in Fig. 3. (Kim et al. 2017) Fig. 4 shows an example of the application to lead-free piezoelectric materials. (Wu et al. 2016) It is known that good properties tend to result from an intimate phase mixture on the nanoscale. Engineering composition to have such phase transitions over the correct temperature range can produce high piezoelectric coefficients, therefore it is of interest to be able to map local ferroelectric displacements cell by cell to track the polarization rotations. Figure 4 reveals a gradual polarization rotation from a rhombohedral phase (displacement along <110>) to a tetragonal phase (displacement along <100>). It is believed that this is indicative of a low polarization anisotropy which leads to low domain wall energy and enhanced piezoelectric coefficient.(Zheng et al. 2017) There are many other examples of locating atomic columns to picometer accuracy.(Borisevich et al. 2010a; Borisevich et al. 2010b; Kimoto et al. 2010; Chang et al. 2011b; Kim et al. 2012; Yankovich et al. 2014; Tang et al. 2015; Dycus et al. 2015; He et al. 2015; Tang et al. 2016)

Besides accurate location of atomic column positions, much progress has also been made in quantifying crystal thickness down to the single unit cell level, based on column intensity, often referred to as atom counting(Ortalan et al. 2010; Katz-Boon et al. 2013; Van Aert et al. 2013; De Backer et al. 2013; Martinez et al. 2014; Jones et al. 2014; Sang et al. 2014; De Backer et al.

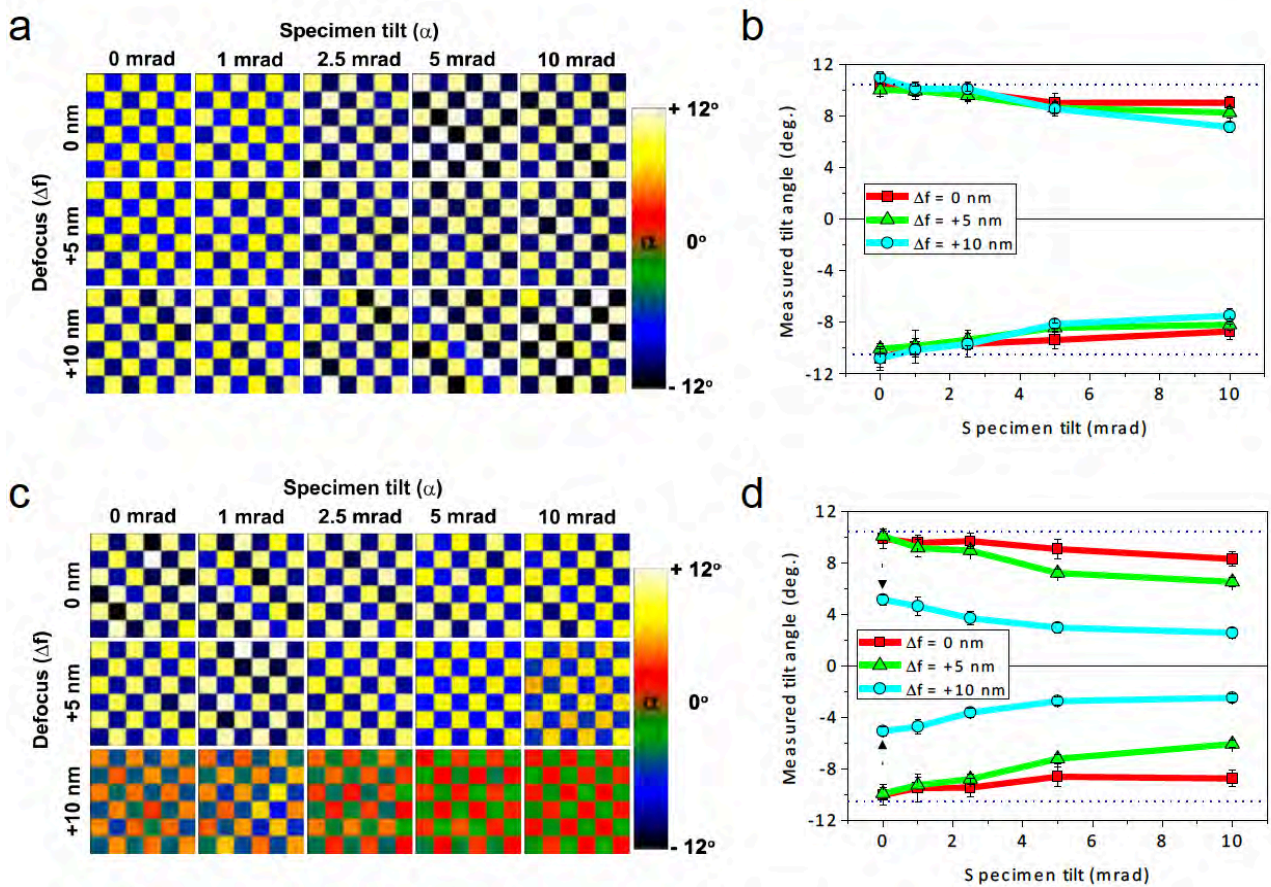

Fig. 3. Effect of specimen tilt in measuring oxygen octahedral tilts in BFO with 10.7 nm thickness. Oxygen octahedral tilt maps as a function of specimen tilt and defocus calculated from simulated (a) BF and (c) ABF images, respectively. (b, d) Line profiles of octahedral tilts averaged over vertical rows of the tilt maps shown in a and c, respectively. The dotted lines in the graphs indicate the tilt value (± 10.5°) of bulk BFO for the $[110]_{pc}$ projection. (Simulations for 100 kV, spherical aberration coefficient -0.037 mm, probe semiangle 31 mrad, BF and ABF collection angles 0 - 1 and 8 - 20 mrad, respectively. Reproduced from Kim et al. 2017. Copyright 2017 with permission from Elsevier.

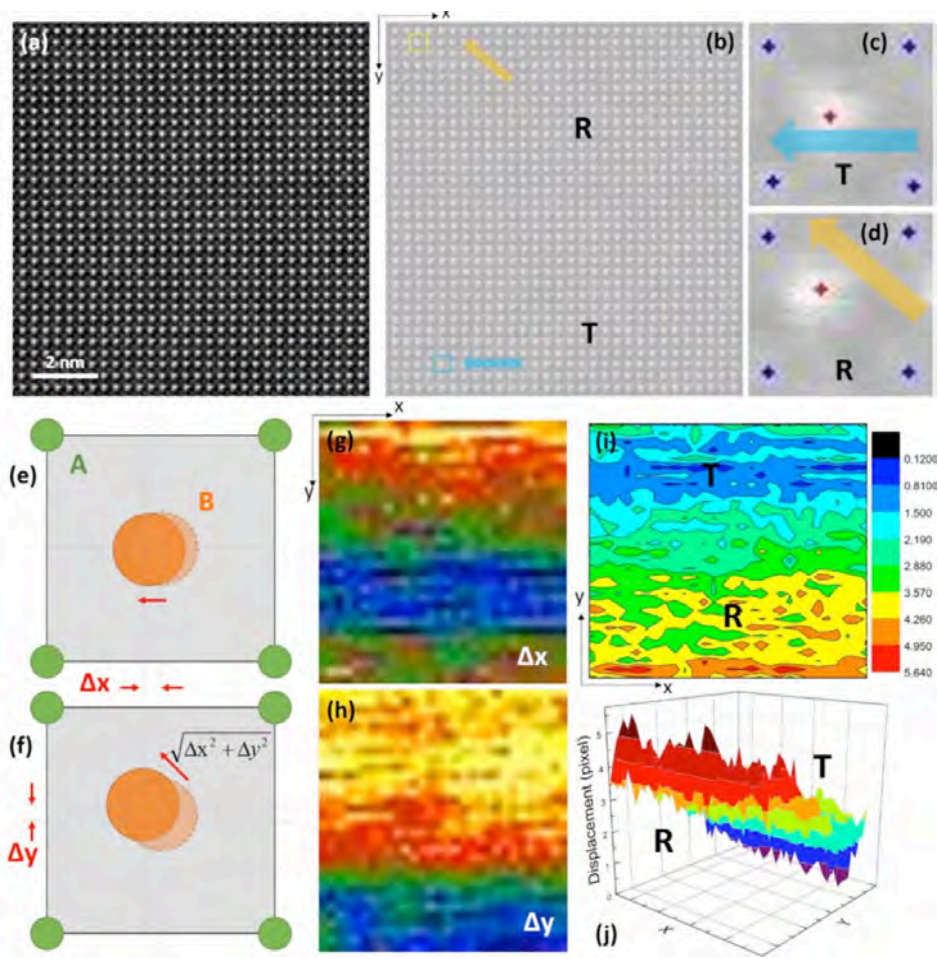

Fig. 4. Local symmetry inside nanodomains of an alkali niobate lead-free ceramic. (a) STEM HAADF image at a domain boundary; (b) Peak finding on (a), revealing rhombohedral (R) and tetragonal (T) regions; (c) Enlarged image of the region in (a) within the blue box, showing T symmetry; (d) Enlarged image of the region in (a) within the yellow box, showing R symmetry; (e,f) Schematics showing projected atom displacements for T and R symmetry; (g,h) Displacements along x and y axes; the region exhibiting displacement along only one axis reflects T symmetry, while the region exhibiting displacement along both x and y axes reflects R symmetry. (i,j) 2D and 3D images showing displacement along the diagonal direction, reflecting different regions with T and R symmetries. Reproduced from Wu et al. 2016. Copyright 2016 American Chemical Society.

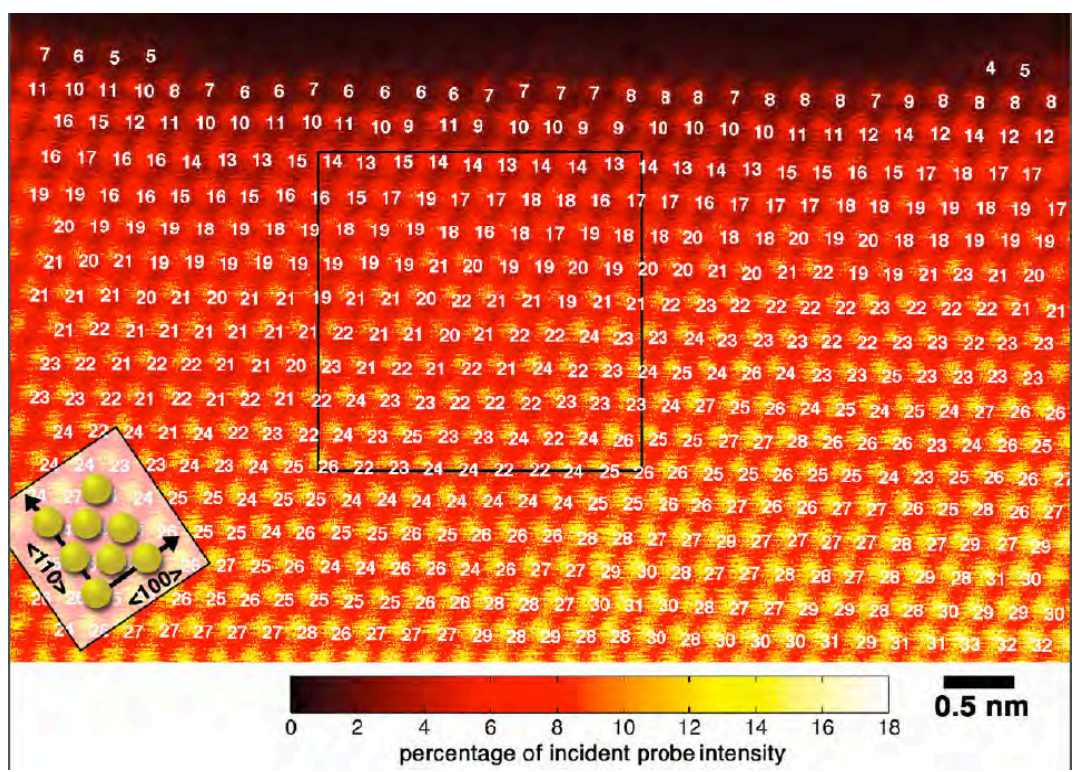

Fig.5. HAADF-STEM image of a wedge-shaped gold film viewed along ⟨110⟩. The intensity maxima correspond to gold atom columns and the white labels near the lower right of each atom column indicate the number of atoms contained in that column. The black box outlines the region from which the PACBED pattern shown in Figure 3 was obtained. The image intensities are shown on an absolute scale relative to the incident beam intensity (see scale bar). Reproduced from LeBeau et al. 2010. Copyright 2010 American Chemical Society.

2015). Figure 5 shows an example for a gold crystal.(LeBeau et al. 2010) It is even possible to locate an impurity atom in depth with single unit cell accuracy by accurate fitting of image intensity to simulations.(Hwang et al. 2013; Ishikawa et al. 2014a) Use of multiple annular detectors provides additional information, (Zhang et al. 2015) and recently it has even been possible to locate vacancies using such quantitative techniques.(Kim et al. 2016)

More complex detector geometries are also becoming popular, particularly a segmented detector which allows a differential phase contrast (DPC) mode. (Shibata et al. 2012; Müller et al. 2014; Lazić et al. 2016) Forming the difference signal between opposite segments provides a measure of beam deflection, which can be used to form an image of electric or magnetic fields. Figure 6 shows the imaging of atomic electric fields in $SrTiO_3$, where the fields are seen to point radially outwards from the centers of all the atomic columns, representing the field between the nuclei and the electrons projected along the viewing direction.

A pixelated detector provides another degree of freedom, in that the entire scattered electron distribution can be recorded for each point in the image. Often referred to as 4D STEM, there are several big advantages although at present the detector readout severely limits the image speed. One major advantage is that the optimum detection angles can be decided after the image scan, not before as required with scintillator detectors. However, a more fundamental advantage is that not just annuli or segments can be used, but more complex detector patterns. This forms the basis of ptychography, where only those regions that give rise to a particular spatial frequency in the image are selected, as shown in Fig. 7. The method produces better signal to noise ratio and higher contrast than bright field imaging since the electrons carrying no signal are excluded. Furthermore, no defocus or aberrations are needed to produce the image so it is ideally complementary to the HAADF image, as shown in Fig. 8.

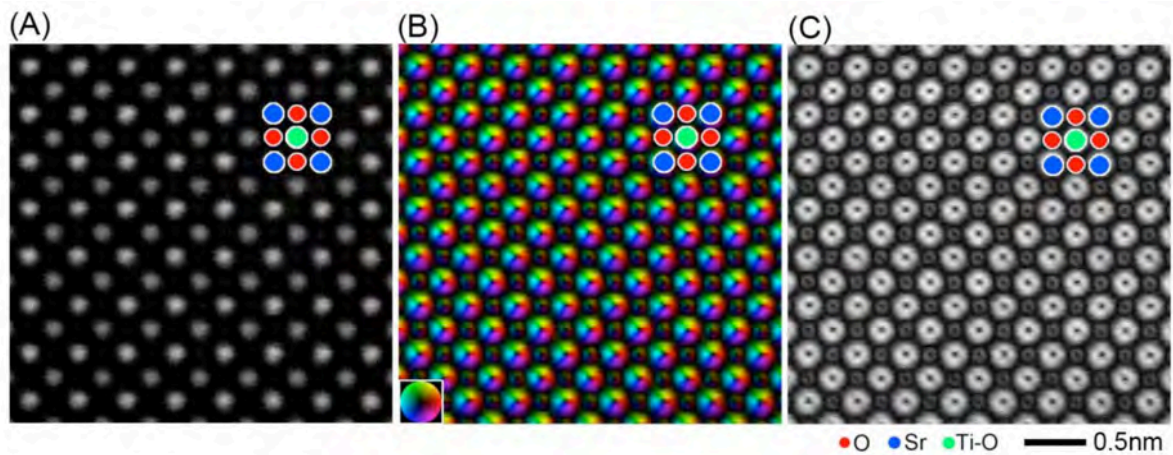

Fig. 6. Simultaneously acquired atomic-resolution STEM images of SrTiO$_3$ [001]. (a) ADF STEM image. (b) Projected electric field vector color map (left side) and electric field strength map (right side) constructed from the segmented-detector STEM images. The inset color wheel indicates how color and shade denote the electric field orientation and strength in the vector color map. It is seen that both heavy and light element columns are sensitively imaged. Intensity dips are clearly visible at the center of each atomic column position. Images taken with a JEOL ARM-300CF operating at 300 kV, adapted from Shibata et al. 2017.

2.2 Analytical Modes

Analytical signals such as EDX and EELS may be orders of magnitude weaker than imaging signals involving scattered electrons, as they typically involve inner shell excitation, which has a decreasing cross section with increasing energy loss. Nevertheless, aberration correction allows much larger currents to be focused into atomic sized probes, as shown in Fig. 9. This fact, combined with major improvements in EDX collection efficiency, have made atomic resolution EDX mapping quite viable, see for example the near-atomically abrupt $SrTiO_3/LaAlO_3$ interface mapped in Fig. 10.

EELS images generally have better statistics because EELS scattering is forward peaked and for edges not too high in energy all the inelastically scattered electrons can enter the spectrometer. Hence it is possible to identify single impurity atoms embedded within a crystal. Figure 11 shows a 0.1 % La-doped film of $CaTiO_3$ grown on $SrTiO_3$ by pulsed laser deposition. Using the La M edge, single La atoms can be located in specific columns of the $CaTiO_3$ crystal.

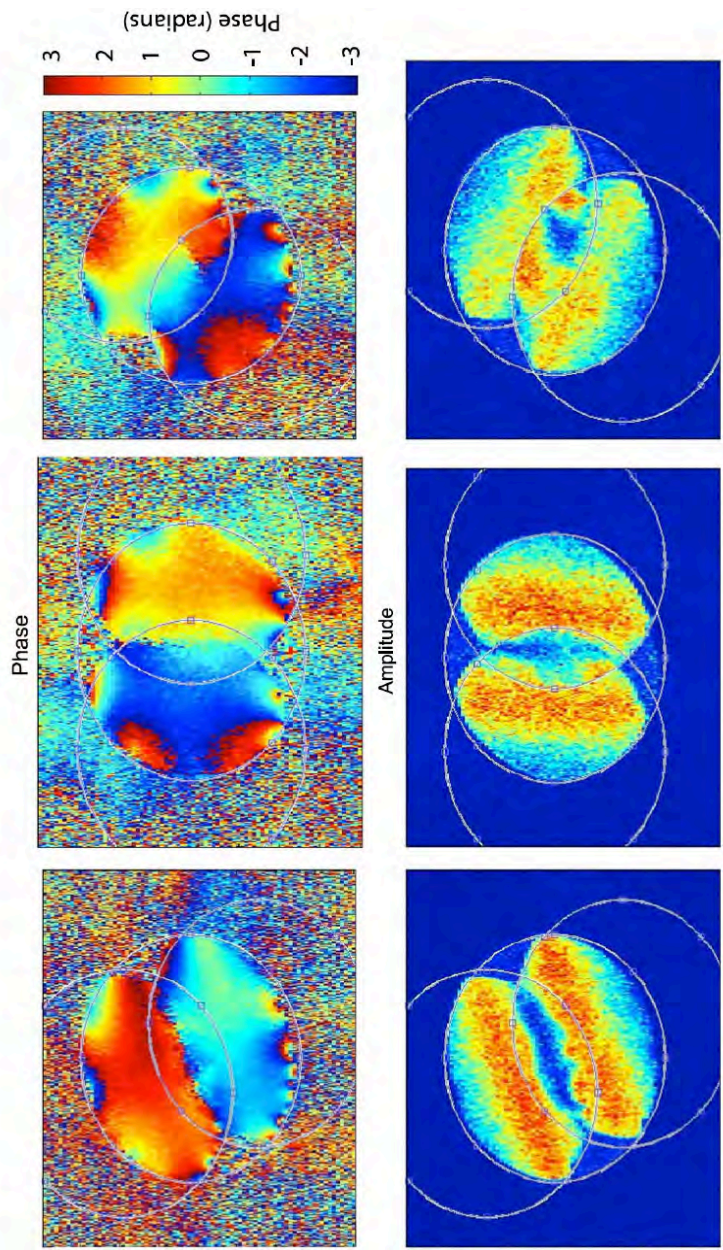

Fig. 7. Ptychographic reconstructions of phase and amplitude components for three spatial frequencies in graphene showing strong contribution from the regions of double overlap. Reproduced from Pennycook et al. 2015. Copyright 2015 with permission from Elsevier.

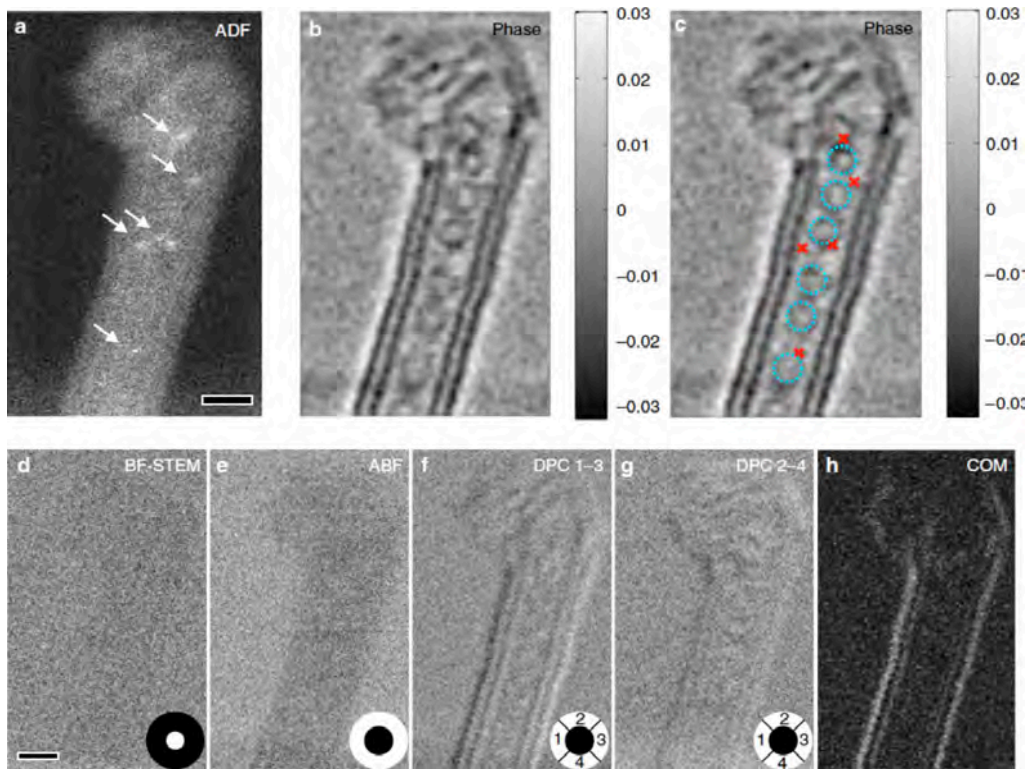

Fig. 8. Simultaneous Z-contrast and phase images of a double-wall carbon nanotube peapod. (a) Incoherent Z-contrast ADF image clearly shows the locations of the single iodine atoms indicated by the arrows. (b) The reconstructed phase image shows the presence of fullerenes inside the nanotube. (c) Annotated phase image with the fullerenes labelled using dotted circles and iodine atoms labelled using cross marks based on their locations in the ADF image. It is clear that the iodine atoms are located close to but outside the fullerenes. For comparison, conventional phase-contrast images including BF, ABF, DPC and the DPC using the centre of mass approach were synthesized from the data and shown in d–h, respectively. The detector area of each imaging method is shown in white colour in d–g. The experiment was performed at an electron probe current of ~2.8 pA, pixel dwell time of 0.25 ms and a dose of ~1.3 $10^4$ e Å $^{-2}$. Scale bar, 1nm; the grey scale of the phase in b is in units of radians. Reproduced from Rutte et al. 2016.

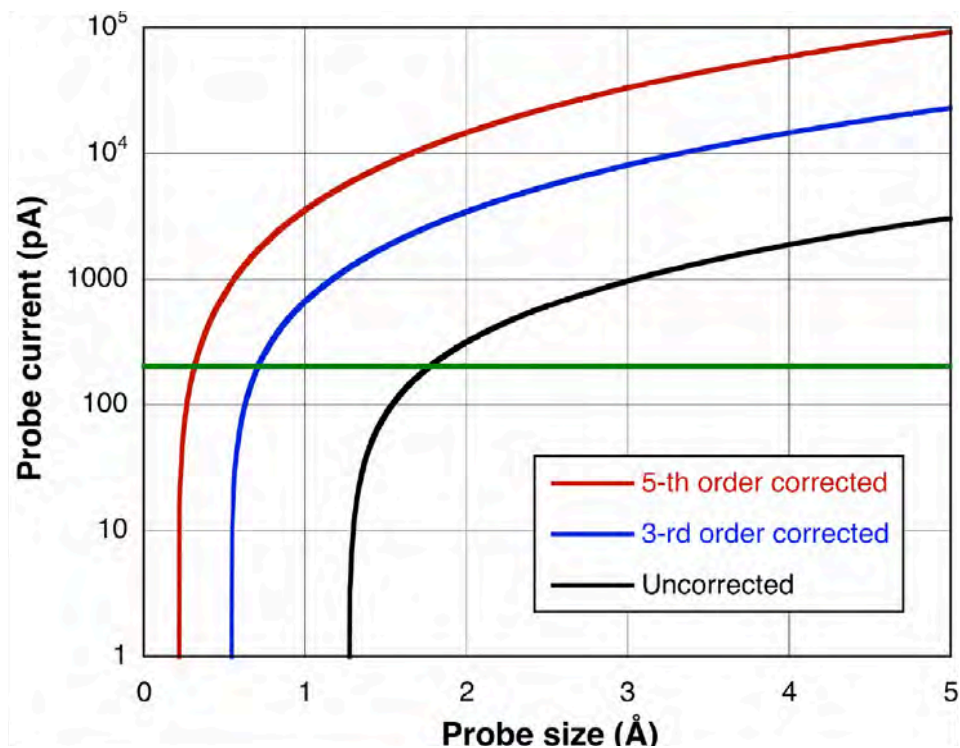

Fig. 9. Variation of probe current with probe size for uncorrected, third- and fifth-order corrected 300 kV microscopes assuming a source brightness of $3\times10^9$ A sr$^{-1}$ cm$^{-2}$. Below the green line the probe is predominantly coherent whereas above it is predominantly incoherent. Adapted from Pennycook 2016 with permission from Springer Nature. Copyright 2016.

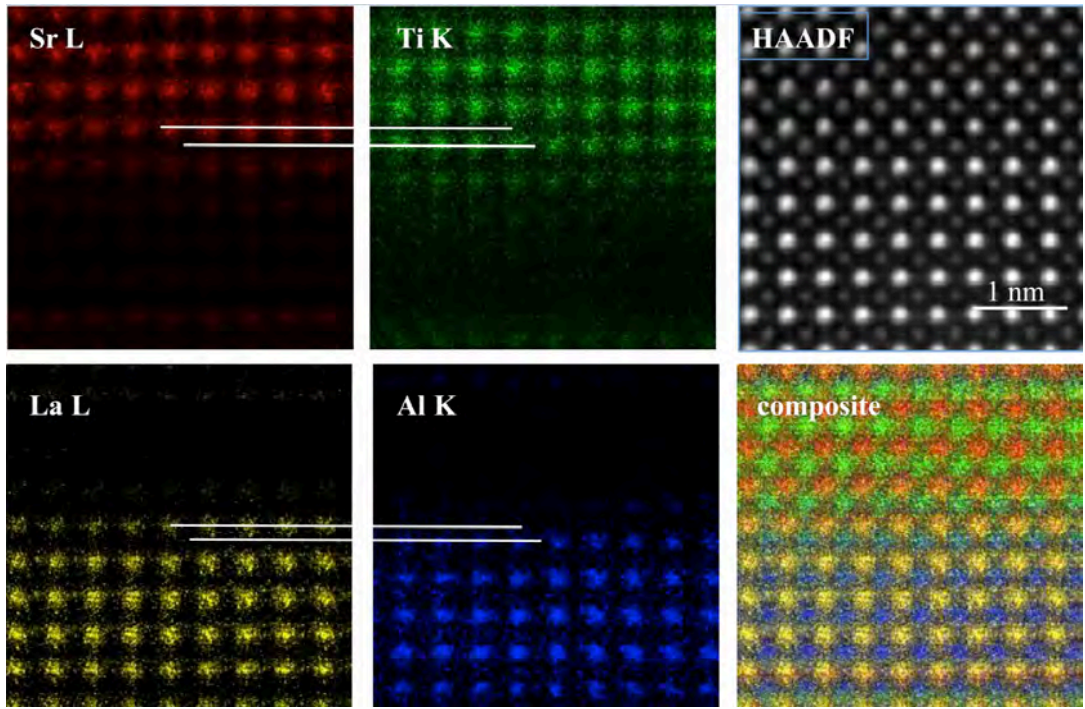

Fig. 10. EDX images of a near atomically abrupt SrTiO$_3$/LaAlO$_3$ interface showing atomic columns identified as labelled, with 10 minutes total acquisition using an Oxford X-Max 100TLE detector on a JEOL ARM200 equipped with ASCOR aberration corrector operated at 200 kV. The composite image includes all edges except O, top right is the HAADF image. Data courtesy Mengsha Li.

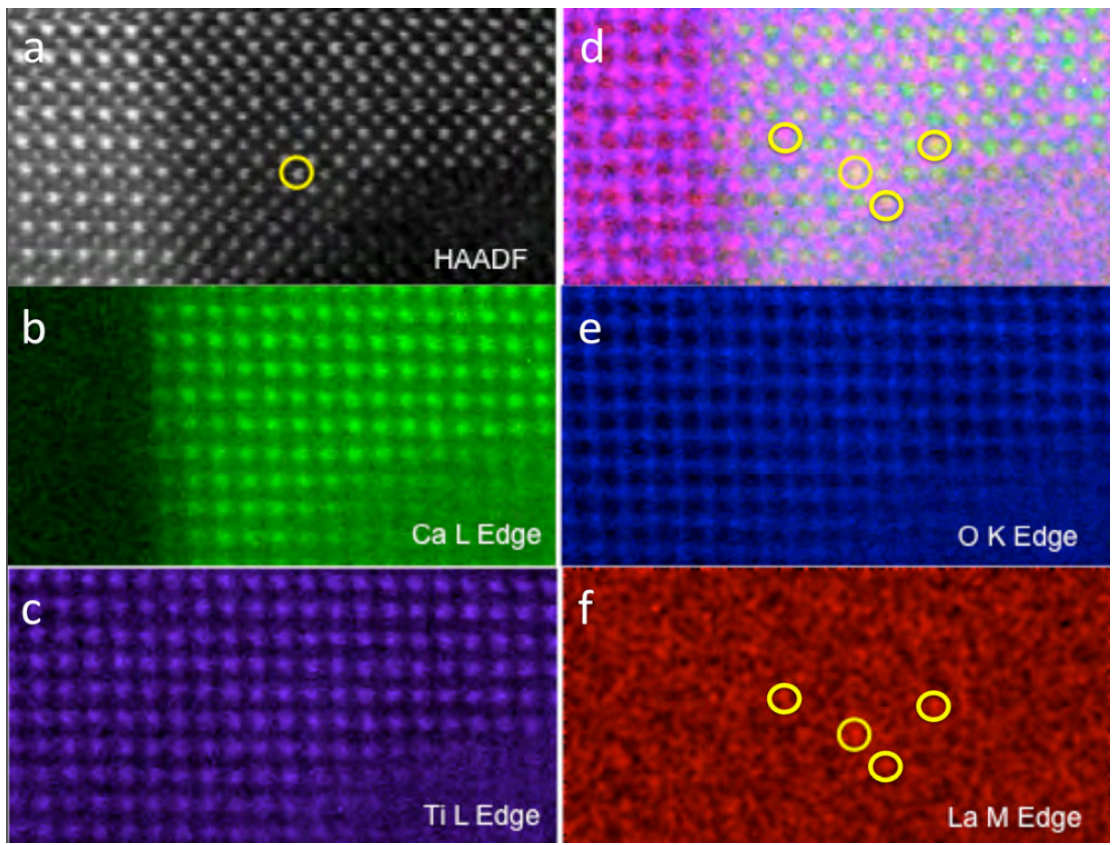

Fig. 11. HAADF image of La-doped $CaTiO_3$ grown on a $SrTiO_3$ substrate with EELS images resolving the Ca, Ti and O columns, and also revealing individual La atoms. The composite color image shows single La atoms in their respective columns. Data obtained with a Gatan Quantum ER on a JEOL ARM 200F equipped with ASCOR aberration corrector operated at 200 kV, recorded and processed by E. Okunishi.

EELS edges also carry information on electronic structure, since the transitions from the core level to the first available empty states depend on atomic valence and environment. This is particularly useful in transition metal oxides since the edge features are linearly related to transition metal valence. Hence oxidation states can be directly extracted from atomic resolution images by comparison to standard spectra, for example the O-K edge of a series of $La_xCa_{1-x}MnO_3$ compounds with varying x, is shown in Fig. 12.(Varela et al. 2009) The first and second peaks can be fitted by Gaussians and parameters such as their ratio or the relative distance between peaks linearly track the Mn oxidation state. This is especially useful for tracking charge transfer across interfaces as shown in Fig. 13, showing EELS data across a $La_{0.7}Ca_{0.3}MnO_3/YBa_2Cu_3O_{7-x}/La_{0.7}Ca_{0.3}MnO_3$ (LCMO/YBCO/LCMO) trilayer.(Varela et al. 2017) By quantifying the fine structure the valence profile can be extracted, and it shows the LCMO layers have a hole concentration slightly higher than bulk while the YBCO layers have a depressed hole concentration, Fig. 14. Hence electrons have transferred from the LCMO into the YBCO, which is consistent with their respective work functions and explains the depressed critical temperatures in the superconductor.(Salafranca et al. 2014) Note also how the normalized prepeak intensity is higher on the Cu-O planes than on the chains, reflecting directly that the holes responsible for superconductivity reside in the planes.

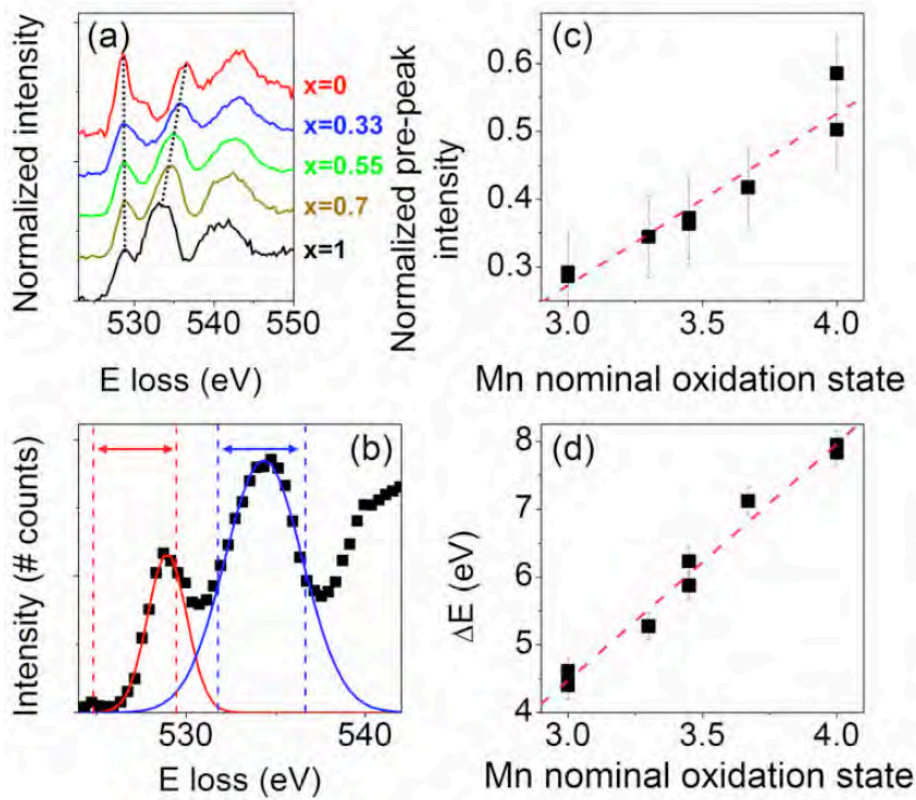

Fig. 12. (a) O K edges for a series of $La_xCa_{1-x}MnO_3$ compounds with x=1, 0.7, 0.55, 0.33 and x=0 from bottom to top. The energy scale has been shifted so the pre-peaks are aligned, and the intensity normalized. The spectra have been displaced vertically for clarity. (b) O K EEL spectrum showing the Gaussian curves used to extract peak intensity and position (pre-peak in red and main peak in blue). (c) Normalized pre-peak intensity versus nominal oxidation state for the series of LCMO samples. The dashed line is a linear fit to the data. (d) Energy separation (calculated as the difference between positions of the second peak and the pre-peak) as a function of the Mn nominal oxidation state for the sample set of samples. Reprinted with permission from Varela et al. 2009. Copyright 2009 by the American Physical Society.

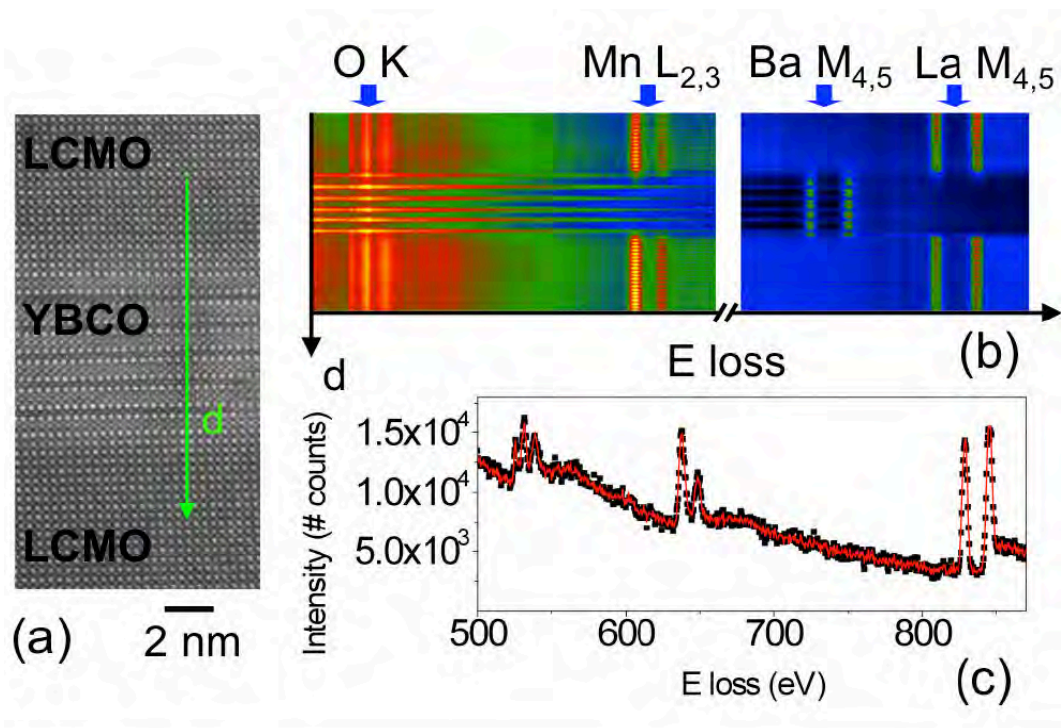

Fig. 13. High resolution Z-contrast image of a LCMO/YBCO/LCMO trilayer, obtained at 100 kV. (b) EELS linescan acquired along the direction marked with an arrow in (a). Principal Component Analysis (PCA) has been used to remove random noise. (c) Sample spectrum extracted from the linescan in (b), acquisition time is 2 seconds per spectrum. The data points are a raw spectrum while the red line is the same dataset after PCA. Reproduced from Varela et al. 2012. Copyright 2012 Oxford University Press.

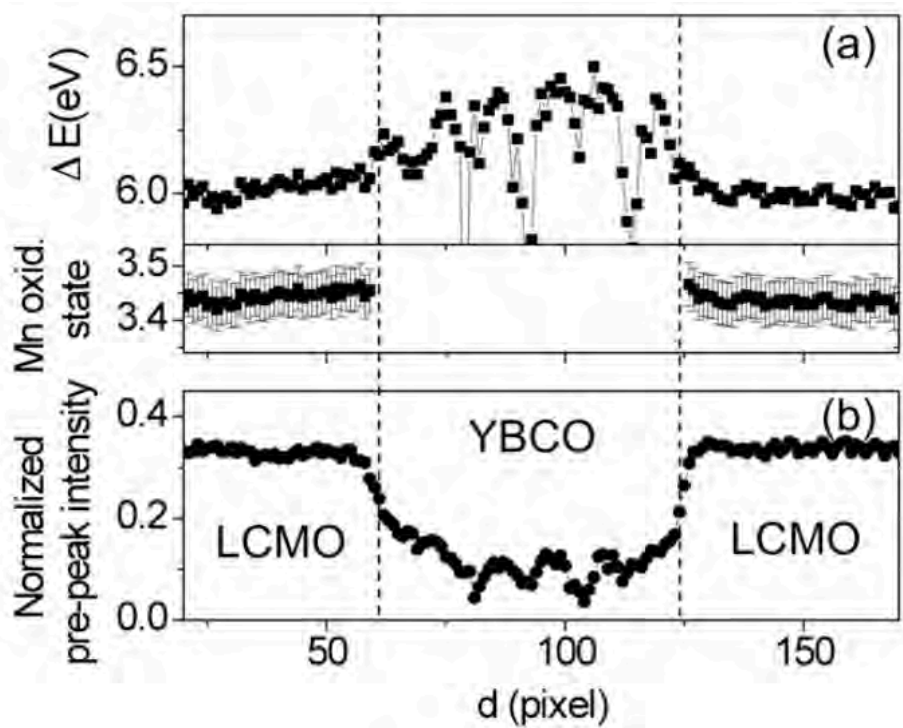

Fig. 14. (a) Top: For the O K edge, peak separation parameter, measured from the linescan in Fig. 13 (b). Bottom: Mn oxidation state in the LCMO layers, derived from the data in (a). (b) For the same linescan, O K edge pre-peak normalized integrated intensity. Reproduced from Varela et al. 2012. Copyright 2012 Oxford University Press.

Another recent major advance is in higher energy resolution, achieved through monochromation, which is reaching into the meV range (Krivanek et al. 2014), opening the door to phonon spectroscopy (Lagos et al. 2017) and band gap mapping.(Lin et al. 2016) Such energy resolution is comparable to that of a synchrotron but the microscope provides much better spatial resolution.

However, especially for low losses, the spatial resolution of EELS images may not be as high as for the HAADF or EDX image because electrons only need to pass close enough to the atom to cause an electronic transition, which can occur some distance away, an effect known as delocalization. Egerton (Egerton 2008) has introduced a measure of delocalization as the diameter containing 50% of the excitations, $d_{50}$. However, it should be noted that this is not the same as image resolution, which is best defined as the full-width-half-maximum of the inelastic image.(Oxley et al. 2016) Because of the delocalization effect, EELS images tend to have long tails more resembling a Lorentzian distribution than a Gaussian. The extended tails reduce image contrast more than they reduce resolution. Note also that delocalization is not a simple function of energy loss but depends on the actual electronic transitions. Recently several examples have been found where low loss images show atomic resolution.(Zhou et al. 2012a; Zhou et al. 2012c; Zhou et al. 2012b) Quantum mechanical simulations show that such contrast arises from specific high momentum transfer transitions, hence there is no violation of the uncertainty principle.(Prange et al. 2012; Oxley et al. 2014; Kapetanakis et al. 2015; Kapetanakis et al. 2016)

3. Dynamics

The high energy electron beam can cause ionization of the sample and direct knock-on events (displacement damage). Ionization damage increases as the beam energy is reduced, but knock-on damage decreases, until it disappears entirely below a certain threshold. Historically such processes have been viewed as undesirable damage events, but now that atomic resolution is possible at lower accelerating voltages there have been many reports of watching atoms move under the beam, allowing insights into atomic motion and the energy landscape of small particles.(Kurasch et al. 2012; Komsa et al. 2012; Komsa et al. 2013; Lin et al. 2014; Yang et al. 2014; Guo et al. 2014; Lehtinen et al. 2015; Jesse et al. 2015) Figure 15 shows a time-sequence of images of a $Si_6$ cluster images at 60 keV beam energy. This is too low to knock atoms out of the cluster, but is sufficient to induce structural changes - one atom is repeatedly seen jumping from left to right. Such studies reveal metastable configurations that would not be seen by simply heating the material.

Another interesting insight into the dynamics of vacancies is revealed in Fig. 16, which shows beam-induced oxygen vacancy ordering in a $LaCoO_3/SrTiO_3$

(LCO/STO) superlattice.(Jang et al. 2017) When O vacancies order into specific lattice planes the strain energy is reduced since the lattice spacing of an entire plane can relax. The plane containing the vacancies expands, causing dark contrast in a Z-contrast image.(Kim et al. 2012) Tracking planar spacings can therefore reveal this ordering process quantitatively. Since the average spacing in the LCO block in Fig. 16 does not change, the images show ordering of pre-existing vacancies rather than generation of new ones.

An example of the beam-induced diffusion of a heavy Ce atom is shown in Fig. 17. Quantification of image intensities matched to image simulations show that the atom jumps between next-neighbor sites inside the crystal, so the beam induced motion is seeing the same diffusion processes that are normally induced thermally. Correlated vacancy/Ce atom motion and interstitial knock-out processes have also been seen.(Ishikawa et al. 2014b) The electron beam can even be used for solid phase epitaxial crystallization of an amorphous material, as demonstrated by writing the letters ORNL in $SrTiO_3$, see Fig. 18 (Jesse et al. 2015). Such beam induced nanolithography should be achievable in three dimensions, due to the nanometer depth of focus of the aberration corrected probe. Recently it has been demonstrated that the STEM probe can be used to "push" single impurity atoms through the graphene lattice.(Dyck et al. 2017) The potential for single atom fabrication with beams and probes is the subject of a recent issue of MRS Bulletin.(Pennycook and Kalinin 2017)

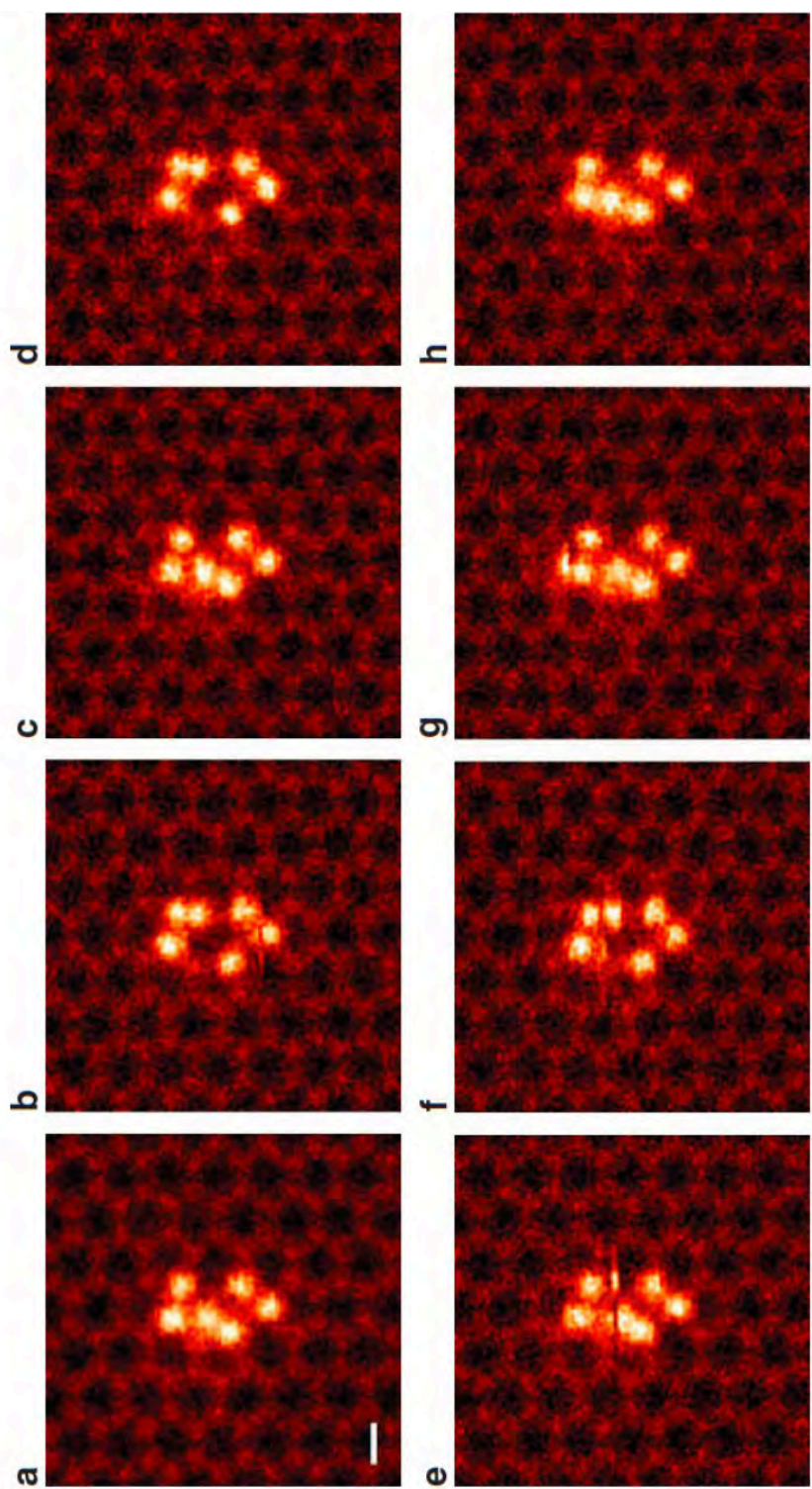

Fig. 15. Sequential STEM Z-contrast images of a $Si_6$ cluster embedded in a graphene pore (a–h). Scale bar, 0.2 nm. Reproduced from Lee et al. 2013.

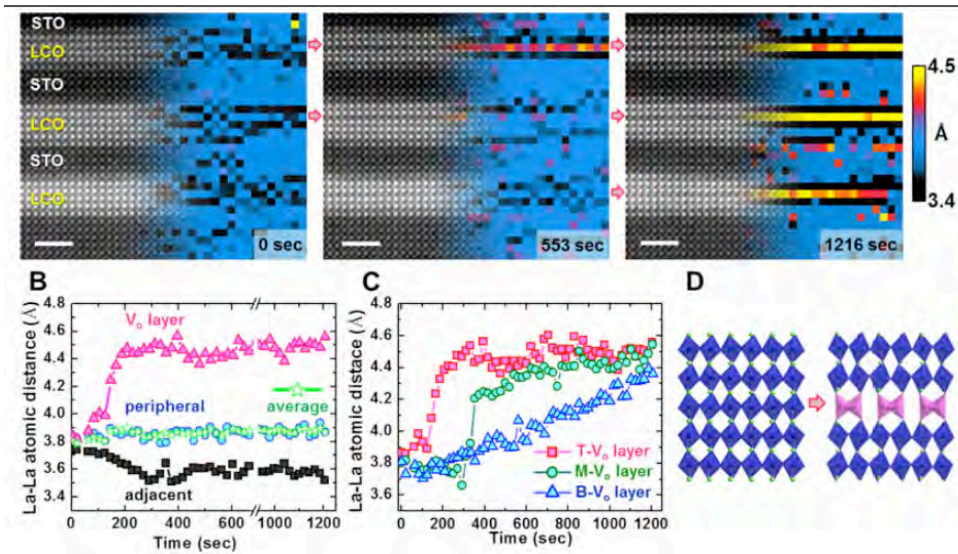

Fig. 16. Beam-induced oxygen vacancy ordering in a LaCoO$_3$/SrTiO$_3$ superlattice grown on a SrTiO$_3$ substrate viewed along the [100] direction. (a) Sequential ADF images overlaid with corresponding maps of the out-of-plane interatomic La-La spacings. (b) Average spacings in the central V$_o$ layer (red curve), the adjacent (black curve) and peripheral (blue curve) layers within the top LCO block as a function of time; green (star) curve gives the overall average, suggesting that the beam primarily induces redistribution of existing vacancies rather than vacancy injection. (c) Comparison of the evolution of interatomic spacings in the oxygen depleted planes of the top (T), middle (M), and bottom (B) LaCoO$_3$ blocks; the differences are attributed to differences in thickness due to the wedge geometry of the sample. (d) Atomic model of the ordering transition; La atoms are shown in green, CoO$_6$ octahedra in blue, and CoO$_4$ tetrahedra in purple. Reproduced from Jang et al. 2017. Copyright 2017 American Chemical Society.

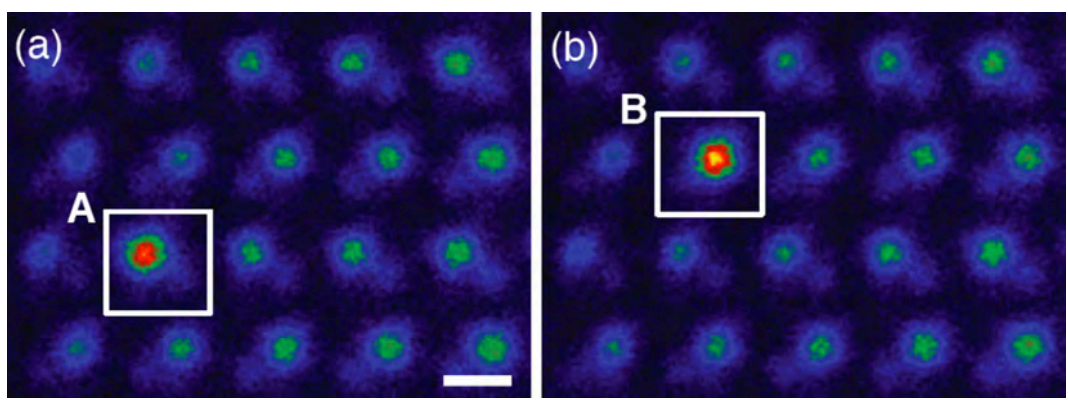

Fig. 17. Atomic-resolution STEM images of Ce hopping from column A to B in AlN. Images are averages over (a) 19 frames before the jump and (b) 19 frames after the jump. Reproduced from Ishikawa et al. 2014a. Copyright 2014 American Chemical Society.

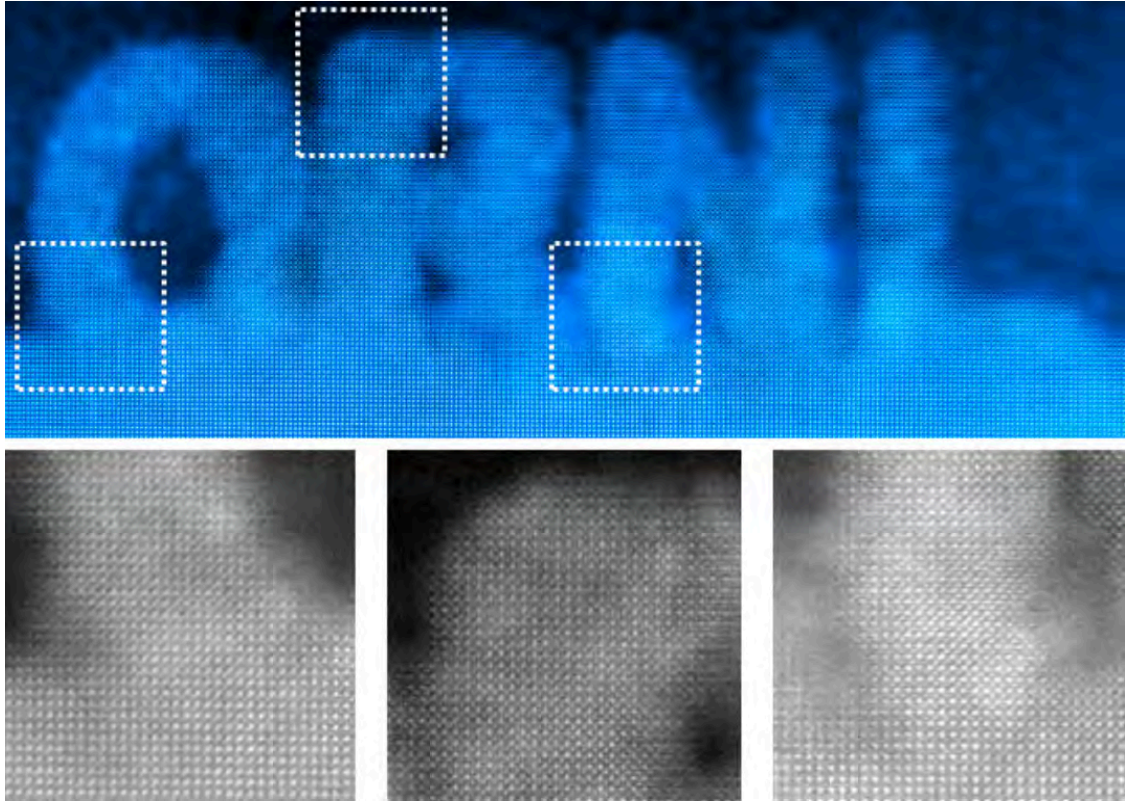

Fig. 18. An example of crystalline oxide sculpting using an arbitrary graphical pattern (in this case, text "ORNL"): Fourier filtered HAADF image of the (a) complete structure and (b-d) magnified raw images of the regions at the top and at the base of the patterned letters. Note the same crystallographic orientation in (b), (c) and (d), highlighting the epitaxial character of the growth. The widths of image (b), (c) and (d) are 12 nm. Reproduced from Jesse et al. 2015.

4. Imaging, analysis and nanofabrication in 3D

Ideally we need atomic scale imaging, analysis and nanofabrication not just in a two-dimensional projection but in three dimensions. In recent years much progress has been made in tomography, combining views in multiple directions to reconstruct 3D structure.(Goris et al. 2013; Bals et al. 2014; Goris et al. 2015; Miao et al. 2016; Bals et al. 2016) Figure 19 shows an example of the atomic scale reconstruction of an FePt nanoparticle revealing the presence of different ordered structures and different degrees of ordering.(Yang et al. 2017) Grain boundaries and even point defects could be detected. Notable progress has also been made in EELS and EDX tomography, although atomic resolution remains far off due to the lower signal levels. (Jarausch et al. 2009; Yedra et al. 2012; Haberfehlner et al. 2014; Collins and Midgley 2017)

The example shown in Fig. 19 required many hours of data collection, which many samples could not withstand. An alternative method is optical sectioning. Aberration correctors achieve higher resolution through higher probe convergence angles. Lateral resolution increases linearly with increasing probe angle, however, depth resolution increases quadratically. So with the latest generation of aberration correctors the depth of focus has reduced to the nanometer scale, and the image comes from a thin section of the sample. Changing the focus gives a series of images at different depths which could also be reconstructed into a 3D image without the need to tilt the sample. At present such optical sectioning has not achieved atomic resolution, but may with future generations of aberration corrector.(Pennycook and Kalinin 2014; Pennycook 2015; Ishikawa et al. 2016) Figure 20 shows a simulated focal series for of Ce atoms substituted in AlN for probe-forming angles of 30, 60 and 100 mrad.(Ishikawa et al. 2015) For the larger probe angles the two dopant atoms can be easily located at 2 and 8 nm depth. Although these simulations assume aberration free conditions, simulations including chromatic aberration and the effects of electron shot noise suggest this approach should work under realistic conditions.(Ishikawa et al. 2016) Optical sectioning can also be applied to analytical signals with nanoscale resolution, avoiding the need for specimen tilting.(Pennycook et al. 2017) Alternatively, it may prove better to combine depth sectioning with discrete tomography (Alania et al. 2017).

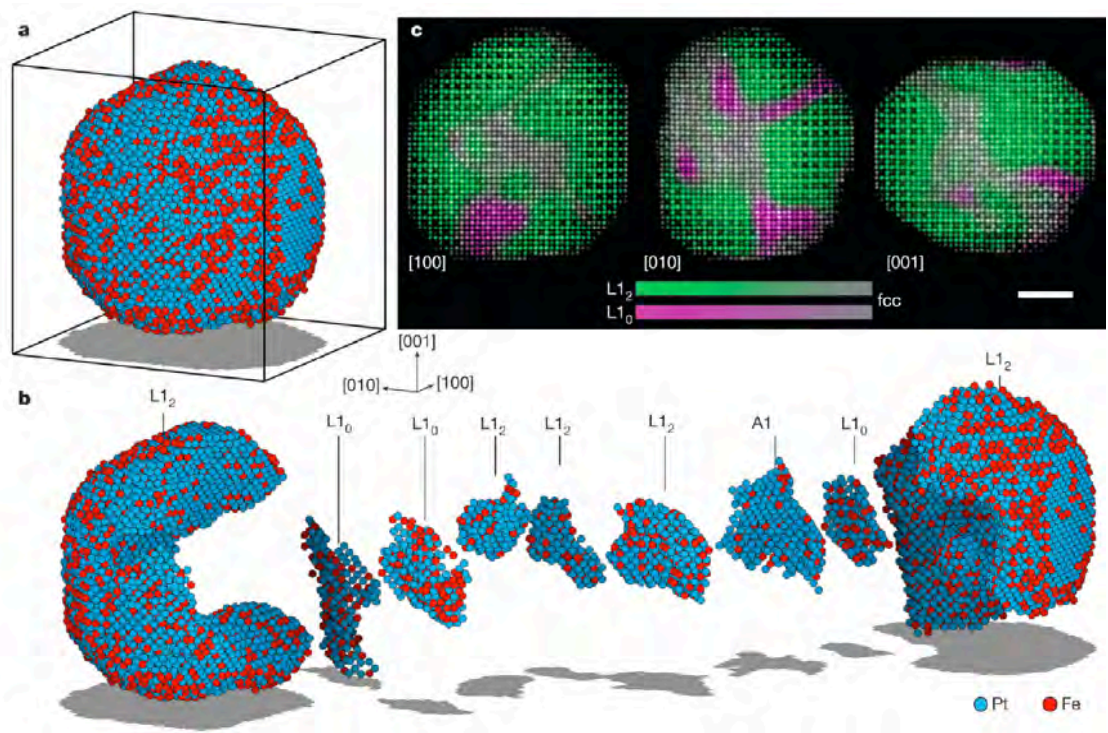

Fig. 19. 3D determination of atomic coordinates, chemical species and grain structure of an FePt nanoparticle. a, Overview of the 3D positions of individual atomic species with Fe atoms in red and Pt atoms in blue. b, The nanoparticle consists of two large L12 grains, three small L12 grains, three small L10 grains and a Pt-rich A1 grain. c, Multislice image. Reproduced from Yang et al. 2017, by permission from Springer Nature. Copyright 2017.

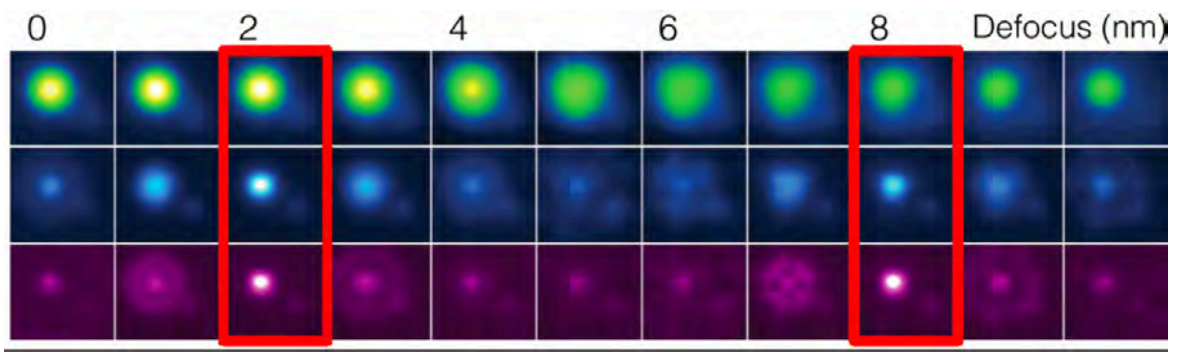

Fig. 20. Simulated focal series of images of Ce atoms substituted in AlN for probe-forming angles of 30 (upper), 60 (centre) and 100 (lower) mrad at 300 kV accelerating voltage, assuming aberration free conditions. Ce atom locations are indicated by red rectangles. Reproduced from Ishikawa et al. 2015. Copyright 2014, with permission from Elsevier.

5. Summary and outlook

STEM has developed dramatically in recent years, thanks to the development of aberration correctors which have allowed the advantages of multiple, simultaneous imaging and spectroscopic modes to be exploited with high sensitivity and precision. There are advantages also for *in situ* and *operandi* studies, since STEM allows good control of dose rate and illumination area.(Chang et al. 2011a; Jungjohann et al. 2012; Mehdi et al. 2015; Wang et al. 2016b) There are also major developments in mathematical image reconstruction techniques, learning from other fields such as computer vision, which are pushing towards lower dose imaging(Stevens et al. 2014; Meyer et al.

2014; Kovarik et al. 2016; Voyles 2017). It is certainly an exciting and rewarding time to be exploring the atomic world.

Acknowledgements: All authors are grateful to their collaborators on the work cited here. SJP is grateful to the National University of Singapore for funding. CJL is supported by a Lee Kuan Yew Fellowship. Research at Universidad Complutense is sponsored by the European Research Council PoC-2016 POLAR-EM and Spanish MINECO-FEDER MAT2015-66888-C3-3-R. YMK was supported by the Institute for Basic Science (IBS-R011-D1) and Creative Materials Discovery Program through the NRF (National Research Foundation of Korea) grant (NRF-2015M3D1A1070672). J.H. Jang is supported by the Korea Basic Science Institute grant (T37210).

Competing Interests

None of the authors have any competing interests in the manuscript.

Figure Captions

Fig. 1. Schematic showing the multiple imaging and analytical signals available in STEM. Top right is a HAADF image of $BiFeO_3$; the brightest atom columns are Bi and the less bright ones Fe, which are seen displaced slightly to the right due to the ferroelectric polarization. The centre image is a bright field image showing O columns, below is a color composite EELS image of $LaMnO_3$ in which the Mn is shown in red, O in green, showing displacements due to octahedral rotations, and La in blue. Adapted with permission from Borisevich et al. 2010b, copyrighted by the American Physical Society, and Varela et al. 2012.

Fig. 2. Measurement of oxygen octahedral tilts in BFO using HAADF and BF STEM images. (a) Schematic of the BFO structure observed along the $[110]_{pc}$ direction. (b, c) Simultaneously acquired HAADF and BF STEM images overlaid with peak finding results for the atomic columns. Bi (red) and Fe (blue) atomic columns can be defined from the HAADF STEM image, (b) and oxygen columns from the BF STEM image, (c). (d) A checkerboard pattern generated from the measurements of the projected tilt angles ($\pm \theta$) of oxygen octahedra. Reproduced from Kim et al. 2017. Copyright 2017 with permission from Elsevier.

Fig. 3. Effect of specimen tilt in measuring oxygen octahedral tilts in BFO with 10.7 nm thickness. Oxygen octahedral tilt maps as a function of specimen tilt and defocus calculated from simulated (a) BF and (c) ABF images, respectively. (b, d) Line profiles of octahedral tilts averaged over vertical rows of the tilt maps shown in a and c, respectively. The dotted lines in the graphs indicate the tilt value ($\pm 10.5°$) of bulk BFO for the $[110]_{pc}$ projection. (Simulations for 100 kV, spherical aberration coefficient -0.037 mm, probe semiangle 31 mrad, BF and

ABF collection angles 0 - 1 and 8 - 20 mrad, respectively. Reproduced from Kim et al. 2017. Copyright 2017 with permission from Elsevier.

Fig. 4. Local symmetry inside nanodomains of an alkali niobate lead-free ceramic. (a) STEM HAADF image at a domain boundary; (b) Peak finding on (a), revealing rhombohedral (R) and tetragonal (T) regions; (c) Enlarged image of the region in (a) within the blue box, showing T symmetry; (d) Enlarged image of the region in (a) within the yellow box, showing R symmetry; (e,f) Schematics showing projected atom displacements for T and R symmetry; (g,h) Displacements along x and y axes; the region exhibiting displacement along only one axis reflects T symmetry, while the region exhibiting displacement along both x and y axes reflects R symmetry. (i,j) 2D and 3D images showing displacement along the diagonal direction, reflecting different regions with T and R symmetries. Reproduced from Wu et al. 2016. Copyright 2016 American Chemical Society.

Fig.5. HAADF-STEM image of a wedge-shaped gold film viewed along ⟨110⟩. The intensity maxima correspond to gold atom columns and the white labels near the lower right of each atom column indicate the number of atoms contained in that column. The black box outlines the region from which the PACBED pattern shown in Figure 3 was obtained. The image intensities are shown on an absolute scale relative to the incident beam intensity (see scale

bar). Reproduced from LeBeau et al. 2010. Copyright 2010 American Chemical Society.

Fig. 6. Simultaneously acquired atomic-resolution STEM images of $SrTiO_3$ [001]. (a) ADF STEM image. (b) Projected electric field vector color map (left side) and electric field strength map (right side) constructed from the segmented-detector STEM images. The inset color wheel indicates how color and shade denote the electric field orientation and strength in the vector color map. It is seen that both heavy and light element columns are sensitively imaged. Intensity dips are clearly visible at the center of each atomic column position. Images taken with a JEOL ARM-300CF operating at 300 kV, adapted from Shibata et al. 2017.

Fig. 7. Ptychographic reconstructions of phase and amplitude components for three spatial frequencies in graphene showing strong contribution from the regions of double overlap. Reproduced from Pennycook et al. 2015. Copyright 2015 with permission from Elsevier.

Fig. 8. Simultaneous Z-contrast and phase images of a double-wall carbon nanotube peapod. (a) Incoherent Z-contrast ADF image clearly shows the locations of the single iodine atoms indicated by the arrows. (b) The reconstructed phase image shows the presence of fullerenes inside the nanotube. (c) Annotated phase image with the fullerenes labelled using dotted

circles and iodine atoms labelled using cross marks based on their locations in the ADF image. It is clear that the iodine atoms are located close to but outside the fullerenes. For comparison, conventional phase-contrast images including BF, ABF, DPC and the DPC using the centre of mass approach were synthesized from the data and shown in d–h, respectively. The detector area of each imaging method is shown in white colour in d–g. The experiment was performed at an electron probe current of ~2.8 pA, pixel dwell time of 0.25 ms and a dose of ~1.3 $10^4$e Å$^{-2}$. Scale bar, 1nm; the grey scale of the phase in b is in units of radians. Reproduced from Rutte et al. 2016.

Fig. 9. Variation of probe current with probe size for uncorrected, third- and fifth-order corrected 300 kV microscopes assuming a source brightness of $3\times10^9$ A sr$^{-1}$ cm$^{-2}$. Below the green line the probe is predominantly coherent whereas above it is predominantly incoherent. Adapted from Pennycook 2016 with permission from Springer Nature. Copyright 2016.

Fig. 10. EDX images of a near atomically abrupt $SrTiO_3$/$LaAlO_3$ interface showing atomic columns identified as labelled, with 10 minutes total acquisition using an Oxford X-Max 100TLE detector on a JEOL ARM200 equipped with ASCOR aberration corrector operated at 200 kV. The composite image includes all edges except O, top right is the HAADF image. Data courtesy Mengsha Li.

Fig. 11. HAADF image of La-doped $CaTiO_3$ grown on a $SrTiO_3$ substrate with EELS images resolving the Ca, Ti and O columns, and also revealing individual La atoms. The composite color image shows single La atoms in their respective columns. Data obtained with a Gatan Quantum ER on a JEOL ARM 200F equipped with ASCOR aberration corrector operated at 200 kV, recorded and processed by E. Okunishi.

Fig. 12. (a) O K edges for a series of $La_xCa_{1-x}MnO_3$ compounds with x=1, 0.7, 0.55, 0.33 and x=0 from bottom to top. The energy scale has been shifted so the pre-peaks are aligned, and the intensity normalized. The spectra have been displaced vertically for clarity. (b) O K EEL spectrum showing the Gaussian curves used to extract peak intensity and position (pre-peak in red and main peak in blue). (c) Normalized pre-peak intensity versus nominal oxidation state for the series of LCMO samples. The dashed line is a linear fit to the data. (d) Energy separation (calculated as the difference between positions of the second peak and the pre-peak) as a function of the Mn nominal oxidation state for the sample set of samples. Reprinted with permission from Varela et al. 2009. Copyright 2009 by the American Physical Society.

Fig. 13. High resolution Z-contrast image of a LCMO/YBCO/LCMO trilayer, obtained at 100 kV. (b) EELS linescan acquired along the direction marked with an arrow in (a). Principal Component Analysis (PCA) has been used to remove random noise. (c) Sample spectrum extracted from the linescan in (b),

acquisition time is 2 seconds per spectrum. The data points are a raw spectrum while the red line is the same dataset after PCA. Reproduced from Varela et al. 2012. Copyright 2012 Oxford University Press.

Fig. 14. (a) Top: For the O K edge, peak separation parameter, measured from the linescan in Fig. 13 (b). Bottom: Mn oxidation state in the LCMO layers, derived from the data in (a). (b) For the same linescan, O K edge pre-peak normalized integrated intensity. Reproduced from Varela et al. 2012. Copyright 2012 Oxford University Press.

Fig. 15. Sequential STEM Z-contrast images of a $Si_6$ cluster embedded in a graphene pore (a–h). Scale bar, 0.2 nm. Reproduced from Lee et al. 2013.

Fig. 16. Beam-induced oxygen vacancy ordering in a $LaCoO_3/SrTiO_3$ superlattice grown on a $SrTiO_3$ substrate viewed along the [100] direction. (a) Sequential ADF images overlaid with corresponding maps of the out-of-plane interatomic La-La spacings. (b) Average spacings in the central $V_o$ layer (red curve), the adjacent (black curve) and peripheral (blue curve) layers within the top LCO block as a function of time; green (star) curve gives the overall average, suggesting that the beam primarily induces redistribution of existing vacancies rather than vacancy injection. (c) Comparison of the evolution of interatomic spacings in the oxygen depleted planes of the top (T), middle (M), and bottom (B) $LaCoO_3$ blocks; the differences are attributed to differences in thickness due to the wedge geometry of the sample. (d) Atomic model of the ordering transition; La atoms are shown in green, $CoO_6$ octahedra in blue, and $CoO_4$ tetrahedra in purple. Reproduced from Jang et al. 2017. Copyright 2017 American Chemical Society.

Fig. 17. Atomic-resolution STEM images of Ce hopping from column A to B in AlN. Images are averages over (a) 19 frames before the jump and (b) 19 frames after the jump. Reproduced from Ishikawa et al. 2014a. Copyright 2014 American Chemical Society.

Fig. 18. An example of crystalline oxide sculpting using an arbitrary graphical pattern (in this case, text "ORNL"): Fourier filtered HAADF image of the (a) complete structure and (b-d) magnified raw images of the regions at the top and at the base of the patterned letters. Note the same crystallographic orientation in (b), (c) and (d), highlighting the epitaxial character of the growth. The widths of image (b), (c) and (d) are 12 nm. Reproduced from Jesse et al. 2015.

Fig. 19. 3D determination of atomic coordinates, chemical species and grain structure of an FePt nanoparticle. a, Overview of the 3D positions of individual atomic species with Fe atoms in red and Pt atoms in blue. **b**, The nanoparticle consists of two large L12 grains, three small L12 grains, three small L10 grains and a Pt-rich A1 grain. **c**, Multislice image. Reproduced from Yang et al. 2017, by permission from Springer Nature. Copyright 2017.

Fig. 20. Simulated focal series of images of Ce atoms substituted in AlN for probe-forming angles of 30 (upper), 60 (centre) and 100 (lower) mrad at 300 kV accelerating voltage, assuming aberration free conditions. Ce atom locations are indicated by red rectangles. Reproduced from Ishikawa et al. 2015. Copyright 2014, with permission from Elsevier.

References

Alania M, Altantzis T, De Backer A, et al (2017) Depth sectioning combined with atom-counting in HAADF STEM to retrieve the 3D atomic structure. Ultramicroscopy 177:36–42. doi: 10.1016/j.ultramic.2016.11.002

Bals S, Goris B, Altantzis T, et al (2014) Seeing and measuring in 3D with electrons. Comptes Rendus Physique 15:140–150. doi: 10.1016/j.crhy.2013.09.015

Bals S, Goris B, De Backer A, et al (2016) Atomic resolution electron tomography. MRS Bull 41:525–530. doi: 10.1557/mrs.2016.138

Borisevich A, Ovchinnikov OS, Chang HJ, et al (2010a) Mapping Octahedral Tilts and Polarization Across a Domain Wall in BiFeO$_3$ from Z-Contrast Scanning Transmission Electron Microscopy Image Atomic Column Shape Analysis. ACS Nano 4:6071–6079. doi: 10.1021/nn1011539

Borisevich AY, Chang HJ, Huijben M, et al (2010b) Suppression of Octahedral Tilts and Associated Changes in Electronic Properties at Epitaxial Oxide Heterostructure Interfaces. Phys Rev Lett 105:087204. doi: 10.1103/PhysRevLett.105.087204

Chang H, Kalinin SV, Yang S, et al (2011a) Watching domains grow: In-situ studies of polarization switching by combined scanning probe and scanning transmission electron microscopy. Journal Of Applied Physics 110:052014–7. doi: 10.1063/1.3623779

Chang HJ, Chang HJ, Kalinin SV, et al (2011b) Atomically Resolved Mapping of Polarization and Electric Fields Across Ferroelectric/Oxide Interfaces by Z-contrast Imaging. Adv Mater 2474–2479. doi: 10.1002/adma.201004641

Collins SM, Midgley PA (2017) Progress and opportunities in EELS and EDS tomography. 1–9. doi: 10.1016/j.ultramic.2017.01.003

Crewe AV (1966) Scanning Electron Microscopes - Is High Resolution Possible? Science 154:729–738.

Crewe AV, Wall J, Langmore J (1970) Visibility of single atoms. Science 168:1338–1340.

De Backer A, De wael A, Gonnissen J, Van Aert S (2015) Optimal experimental design for nano-particle atom-counting from high-resolution STEM images. Ultramicroscopy 151:46–55. doi: 10.1016/j.ultramic.2014.10.015

De Backer A, Martinez GT, Rosenauer A, Van Aert S (2013) Atom counting in HAADF STEM using a statistical model-based approach: Methodology, possibilities, and inherent limitations. Ultramicroscopy 134:23–33. doi: 10.1016/j.ultramic.2013.05.003

Dyck O, Kim S, Kalinin SV, Jesse S (2017) Placing single atoms in graphene with a scanning transmission electron microscope. Applied Physics Letters 111:113104–6. doi: 10.1063/1.4998599

Dycus JH, Harris JS, Sang X, et al (2015) Accurate Nanoscale Crystallography in Real-Space Using Scanning Transmission Electron Microscopy. Microscopy and Analysis 21:946–952. doi: 10.1017/S1431927615013732

Egerton RF (2008) Electron energy-loss spectroscopy in the TEM. Rep Prog Phys 72:016502. doi: 10.1088/0034-4885/72/1/016502

Findlay SD, Shibata N, Sawada H, et al (2010) Dynamics of annular bright field imaging in scanning transmission electron microscopy. Ultramicroscopy 110:903–923. doi: 10.1016/j.ultramic.2010.04.004

Gazquez J, Sanchez-Santolino G, Biškup N, et al (2017) Materials Science in Semiconductor Processing. Materials Science in Semiconductor Processing 65:49–63. doi: 10.1016/j.mssp.2016.06.005

Goris B, De Backer A, Van Aert S, et al (2013) Three-Dimensional Elemental Mapping at the Atomic Scale in Bimetallic Nanocrystals. Nano Lett 13:4236–4241. doi: 10.1021/nl401945b

Goris B, De Beenhouwer J, De Backer A, et al (2015) Measuring Lattice Strain in Three Dimensions through Electron Microscopy. Nano Lett 15:6996–7001. doi: 10.1021/acs.nanolett.5b03008

Guo J, Lee J, Contescu CI, et al (2014) Crown ethers in graphene. Nature Communications 5:1–6. doi: 10.1038/ncomms6389

Haberfehlner G, Orthacker A, Albu M, et al (2014) Nanoscale voxel spectroscopy by simultaneous EELS and EDS tomography. Nanoscale 6:14563–14569. doi: 10.1039/C4NR04553J

He Q, Ishikawa R, Lupini AR, et al (2015) Towards 3D Mapping of $BO_6$ Octahedron Rotations at Perovskite Heterointerfaces, Unit Cell by Unit Cell. ACS Nano 9:8412–8419. doi: 10.1021/acsnano.5b03232

Hwang J, Zhang JY, D'Alfonso AJ, et al (2013) Three-Dimensional Imaging of Individual Dopant Atoms in $SrTiO_3$. Phys Rev Lett 111:266101. doi: 10.1103/PhysRevLett.111.266101

Ishikawa R, Lupini AR, Findlay SD, et al (2014a) Three-Dimensional Location of a Single Dopant with Atomic Precision by Aberration-Corrected Scanning Transmission Electron Microscopy. Nano Lett 1903–1908. doi: 10.1021/nl500564b

Ishikawa R, Lupini AR, Hinuma Y, Pennycook SJ (2015) Large-angle illumination STEM: Toward three-dimensional atom-by- atom imaging. Ultramicroscopy 151:122–129. doi: 10.1016/j.ultramic.2014.11.009

Ishikawa R, Mishra R, Lupini AR, et al (2014b) Direct Observation of Dopant Atom Diffusion in a Bulk Semiconductor Crystal Enhanced by a Large Size Mismatch. Phys Rev Lett 113:155501. doi: 10.1103/PhysRevLett.113.155501

Ishikawa R, Okunishi E, Sawada H, et al (2011) Direct imaging of hydrogen-atom columns in a crystal by annular bright-field electron microscopy. Nature Materials 10:278–281. doi: 10.1038/nmat2957

Ishikawa R, Pennycook SJ, Lupini AR, et al (2016) Single atom visibility in STEM optical depth sectioning. Appl Phys Lett 109:163102. doi: 10.1063/1.4965709

Jang JH, Kim Y-M, He Q, et al (2017) In Situ Observation of Oxygen Vacancy Dynamics and Ordering in the Epitaxial $LaCoO_3$ System. ACS Nano 11:6942–6949. doi: 10.1021/acsnano.7b02188

Jarausch K, Thomas P, Leonard DN, et al (2009) Four-dimensional STEM-EELS: Enabling nano-scale chemical tomography. Ultramicroscopy 109:326–337. doi: 10.1016/j.ultramic.2008.12.012

Jesse S, He Q, Lupini AR, et al (2015) Atomic-Level Sculpting of Crystalline Oxides: Toward Bulk Nanofabrication with Single Atomic Plane Precision. Small 11:5895–5900. doi: 10.1002/smll.201502048

Jones L, MacArthur KE, Fauske VT, et al (2014) Rapid Estimation of Catalyst Nanoparticle Morphology and Atomic-Coordination by High-Resolution Z-

Contrast Electron Microscopy. Nano Lett 14:6336–6341. doi: 10.1021/nl502762m

Jungjohann KL, Evans JE, Aguiar JA, et al (2012) Atomic-Scale Imaging and Spectroscopy for In Situ Liquid Scanning Transmission Electron Microscopy. Microscopy and Analysis 18:621–627. doi: 10.1017/S1431927612000104

Kapetanakis MD, Oxley MP, Zhou W, et al (2016) Signatures of distinct impurity configurations in atomic-resolution valence electron-energy-loss spectroscopy: Application to graphene. Phys Rev B 94:155449. doi: 10.1103/PhysRevB.94.155449

Kapetanakis MD, Zhou W, Oxley MP, et al (2015) Low-loss electron energy loss spectroscopy: An atomic-resolution complement to optical spectroscopies and application to graphene. Phys Rev B 92:125147. doi: 10.1103/PhysRevB.92.125147

Katz-Boon H, Rossouw CJ, Dwyer C, Etheridge J (2013) Rapid Measurement of Nanoparticle Thickness Profiles. Ultramicroscopy 124:61–70. doi: 10.1016/j.ultramic.2012.08.009

Kim H, Zhang JY, Raghavan S, Stemmer S (2016) Direct Observation of Sr Vacancies in $SrTiO_3$ by Quantitative Scanning Transmission Electron Microscopy. Phys Rev X 6:041063–7. doi: 10.1103/PhysRevX.6.041063

Kim Y-M, He J, Biegalski MD, et al (2012) Probing oxygen vacancy concentration and homogeneity in solid-oxide fuel-cell cathode materials on the subunit-cell level. Nature Materials 11:1–7. doi: 10.1038/nmat3393

Kim Y-M, Pennycook SJ, Borisevich AY (2017) Quantitative comparison of bright field and annular bright field imaging modes for characterization of oxygen octahedral tilts. Ultramicroscopy 181:1–7. doi: 10.1016/j.ultramic.2017.04.020

Kimoto K, Asaka T, Yu X, et al (2010) Local crystal structure analysis with several picometer precision using scanning transmission electron microscopy. Ultramicroscopy 110:778–782. doi: 10.1016/j.ultramic.2009.11.014

Komsa H-P, Kotakoski J, Kurasch S, et al (2012) Two-Dimensional Transition Metal Dichalcogenides under Electron Irradiation: Defect Production and Doping. Phys Rev Lett 109:035503–5. doi: 10.1103/PhysRevLett.109.035503

Komsa H-P, Kurasch S, Lehtinen O, et al (2013) From point to extended defects in two-dimensional $MoS_2$: Evolution of atomic structure under electron irradiation. Phys Rev B 88:035301–8. doi: 10.1103/PhysRevB.88.035301

Kovarik L, Stevens A, Liyu A, Browning ND (2016) Implementing an accurate and rapid sparse sampling approach for low-dose atomic resolution STEM imaging. Applied Physics Letters 109:164102–6. doi: 10.1063/1.4965720

Krivanek OL, Lovejoy TC, Dellby N, et al (2014) Vibrational spectroscopy in the electron microscope. Nature 514:209–212. doi: 10.1038/nature13870

Kurasch S, Kotakoski J, Lehtinen O, et al (2012) Atom-by-Atom Observation of Grain Boundary Migration in Graphene. Nano Lett 12:3168–3173. doi: 10.1021/nl301141g

Lagos MJ, Trügler A, Hohenester U, Batson PE (2017) Mapping vibrational surface and bulk modes in a single nanocube. Nature 543:529–532. doi: 10.1038/nature21699

Lazić I, Bosch EGT, Lazar S (2016) Phase contrast STEM for thin samples: Integrated differential phase contrast. Ultramicroscopy 160:265–280. doi: 10.1016/j.ultramic.2015.10.011

LeBeau JM, Findlay SD, Allen LJ, Stemmer S (2010) Standardless Atom Counting in Scanning Transmission Electron Microscopy. Nano Lett 10:4405–4408. doi: 10.1021/nl102025s

Lee J, Zhou W, Pennycook SJ, et al (2013) Direct visualization of reversible dynamics in a $Si_6$ cluster embedded in a graphene pore. Nature Communications 4:1650–7. doi: 10.1038/ncomms2671

Lehtinen O, Vats N, Algara-Siller G, et al (2015) Implantation and Atomic-Scale Investigation of Self-Interstitials in Graphene. Nano Lett 15:235–241. doi: 10.1021/nl503453u

Li C, Poplawsky J, Yan Y, Pennycook SJ (2017) Understanding individual defects in CdTe thin-film solar cells via STEM: From atomic structure to electrical activity. Materials Science in Semiconductor Processing 65:64–76. doi: 10.1016/j.mssp.2016.06.017

Lin J, Cretu O, Zhou W, et al (2014) Flexible metallic nanowires with self-adaptive contacts to semiconducting transition-metal dichalcogenide monolayers. Nature nanotechnology 9:436–442. doi: 10.1038/nnano.2014.81

Lin J, Gomez L, de Weerd C, et al (2016) Direct Observation of Band Structure Modifications in Nanocrystals of $CsPbBr_3$ Perovskite. Nano Lett 16:7198–7202. doi: 10.1021/acs.nanolett.6b03552

Martinez GT, Rosenauer A, De Backer A, et al (2014) Quantitative composition determination at the atomic level using model-based high-angle annular dark field scanning transmission electron microscopy. Ultramicroscopy 137:12–19. doi: 10.1016/j.ultramic.2013.11.001

Mehdi BL, Qian J, Nasybulin E, et al (2015) Observation and Quantification of Nanoscale Processes in Lithium Batteries by Operando Electrochemical (S)TEM. Nano Lett 15:2168–2173. doi: 10.1021/acs.nanolett.5b00175

Meyer JC, Kotakoski J, Mangler C (2014) Atomic structure from large-area, low-dose exposures of materials: A new route to circumvent radiation damage. Ultramicroscopy 145:13–21. doi: 10.1016/j.ultramic.2013.11.010

Miao J, Ercius P, Billinge S (2016) Atomic electron tomography: 3D structures without crystals. Science. doi: 10.1126/science.aaf2157

Müller K, Krause FF, Béché A, et al (2014) Atomic electric fields revealed by a quantum mechanical approach to electron picodiffraction. Nature Communications 5:5653. doi: 10.1038/ncomms6653

Ortalan V, Uzun A, Gates BC, Browning ND (2010) Towards full-structure determination of bimetallic nanoparticles with an aberration-corrected electron microscope. Nature nanotechnology 5:843–847. doi: 10.1038/nnano.2010.234

Oxley MP, Kapetanakis MD, Prange MP, et al (2014) Simulation of Probe Position-Dependent Electron Energy-Loss Fine Structure. Microscopy and Analysis 20:784–797. doi: 10.1017/S1431927614000610

Oxley MP, Lupini AR, Pennycook SJ (2016) Ultra-high resolution electron microscopy. Rep Prog Phys 80:026101–65. doi: 10.1088/1361-6633/80/2/026101

Pennycook SJ (2016) Imaging in STEM. In: Carter CB, Williams DB (eds) Transmission Electron Microscopy, 1st edn. Springer, pp 283–342

Pennycook SJ (2015) Fulfilling Feynman's dream: "Make the electron microscope 100 times better"—Are we there yet? MRS Bull 40:71–78. doi: 10.1557/mrs.2014.307

Pennycook SJ, Boatner LA (1988) Chemically sensitive structure-imaging with a scanning transmission electron microscope. Nature 336:565–567.

Pennycook SJ, Jesson DE (1990) High-resolution incoherent imaging of crystals. Phys Rev Lett 64:938–941.

Pennycook SJ, Jesson DE (1991) High-resolution Z-contrast imaging of crystals. Ultramicroscopy 37:14–38.

Pennycook SJ, Kalinin SV (2017) Single atom fabrication with beams and probes.

Pennycook SJ, Kalinin SV (2014) Hasten high resolution. Nature 515:487–488.

Pennycook SJ, Nellist PD (eds) (2011) Scanning Transmission Electron Microscopy. Springer New York, New York, NY

Pennycook TJ, Lupini AR, Yang H, et al (2015) Efficient phase contrast imaging in STEM using a pixelated detector. Part 1: Experimental demonstration at atomic resolution. Ultramicroscopy 151:160–167. doi: 10.1016/j.ultramic.2014.09.013

Pennycook TJ, Yang H, Jones L, et al (2017) 3D elemental mapping with nanometer scale depth resolution via electron optical sectioning. Ultramicroscopy 174:27–34. doi: 10.1016/j.ultramic.2016.12.002

Prange MP, Oxley MP, Varela M, et al (2012) Simulation of Spatially Resolved Electron Energy Loss Near-Edge Structure for Scanning Transmission Electron Microscopy. Phys Rev Lett 109:246101. doi: 10.1103/PhysRevLett.109.246101

Rose H (1974) Phase-contrast in scanning transmission electron microscopy. Optik 39:416–436.

Rutte RN, Jones L, Simson M, et al (2016) Simultaneous atomic-resolution electron ptychography and Z-contrast imaging of light and heavy elements in complex nanostructures. Nature Communications 7:1–8. doi: 10.1038/ncomms12532

Salafranca J, Rincón J, Tornos J, et al (2014) Competition between Covalent Bonding and Charge Transfer at Complex-Oxide Interfaces. Phys Rev Lett 112:196802.

Sang X, Oni AA, LeBeau JM (2014) Atom Column Indexing: Atomic Resolution Image Analysis Through a Matrix Representation. Microscopy and Analysis 20:1764–1771. doi: 10.1017/S1431927614013506

Shibata N, Findlay SD, Kohno Y, et al (2012) Differential phase-contrast microscopy at atomic resolution. Nat Phys 8:611–615. doi: 10.1038/nphys2337

Shibata N, Seki T, Sanchez-Santolino G, et al (2017) Electric field imaging of single atoms. Nature Communications 8:15631. doi: 10.1038/ncomms15631

Stevens A, Yang H, Carin L, et al (2014) The potential for Bayesian compressive sensing to significantly reduce electron dose in high-resolution STEM images. Microscopy 63:41–51. doi: 10.1093/jmicro/dft042

Tang YL, Zhu YL, Hong ZJ, et al (2016) 3D polarization texture of a symmetric 4-fold flux closure domain in strained ferroelectric PbTiO$_3$ films. J Mater Res 32:957–967. doi: 10.1557/jmr.2016.259

Tang YL, Zhu YL, Ma XL, et al (2015) Observation of a periodic array of flux-closure quadrants in strained ferroelectric PbTiO$_3$ films. Science 348:547–551. doi: 10.1126/science.1258289

Van Aert S, De Backer A, Martinez G, et al (2013) Procedure to count atoms with trustworthy single-atom sensitivity. Phys Rev B 87:064107. doi: 10.1103/PhysRevB.87.064107

Varela M, Gazquez J, Pennycook SJ (2012) STEM-EELS imaging of complex oxides and interfaces. MRS Bull 37:29–35. doi: 10.1557/mrs.2011.330

Varela M, Leon C, Santamaria J, Pennycook SJ (2012) Scanning transmission electron microscopy of oxides. In: Tsymbal EY, Dagotto ERA, Eom C-B, Ramesh R (eds) Multifunctional Oxide Heterostructures. Oxford University Press, pp 1–51

Varela M, Oxley MP, Luo W, et al (2009) Atomic-resolution imaging of oxidation states in manganites. Phys Rev B 79:085117. doi: 10.1103/PhysRevB.79.085117

Voyles PM (2017) Informatics and data science in materials microscopy. Current Opinion in Solid State & Materials Science 21:141–158. doi: 10.1016/j.cossms.2016.10.001

Wang Y, Salzberger U, Sigle W, et al (2016a) Ultramicroscopy. Ultramicroscopy 168:46–52. doi: 10.1016/j.ultramic.2016.06.001

Wang Z, Santhanagopalan D, Zhang W, et al (2016b) In Situ STEM-EELS Observation of Nanoscale Interfacial Phenomena in All-Solid-State Batteries. Nano Lett 16:3760–3767. doi: 10.1021/acs.nanolett.6b01119

Wu B, Wu H, Wu J, et al (2016) Giant Piezoelectricity and High Curie Temperature in Nanostructured Alkali Niobate Lead-Free Piezoceramics through Phase Coexistence. J Am Chem Soc 138:15459–15464. doi: 10.1021/jacs.6b09024

Yang Y, Chen C-C, Scott MC, et al (2017) Deciphering chemical order/disorder and material properties at the single-atom level. Nature 542:75–79. doi: 10.1038/nature21042

Yang Z, Yin L, Lee J, et al (2014) Direct Observation of Atomic Dynamics and Silicon Doping at a Topological Defect in Graphene. Angew Chem Int Edit 53:8908–8912. doi: 10.1002/anie.201403382

Yankovich AB, Berkels B, Dahmen W, et al (2014) Picometre-precision analysis of scanning transmission electron microscopy images of platinum nanocatalysts. Nature Communications 5:1–7. doi: 10.1038/ncomms5155

Yedra L, Eljarrat A, Arenal R, et al (2012) EEL spectroscopic tomography Towards a new dimension in nanomaterials analysis. Ultramicroscopy 122:12–18. doi: 10.1016/j.ultramic.2012.07.020

Zhang JY, Hwang J, Isaac BJ, Stemmer S (2015) Variable-angle high-angle annular dark-field imaging: application to three-dimensional dopant atom profiling. Sci Rep 1–10. doi: 10.1038/srep12419

Zheng T, Wu H, Yuan Y, et al (2017) The structural origin of enhanced piezoelectric performance and stability in lead free ceramics. Energy Environ Sci 10:528–537. doi: 10.1039/C6EE03597C

Zhou W, Lee J, Nanda J, et al (2012a) Atomically localized plasmon enhancement in monolayer graphene. Nature nanotechnology 7:161–165. doi: 10.1038/NNANO.2011.252

Zhou W, Oxley MP, Lupini AR, et al (2012b) Single Atom Microscopy. Microscopy and Analysis 18:1342–1354. doi: 10.1017/S1431927612013335

Zhou W, Pennycook SJ, Idrobo J-C (2012c) Localization of inelastic electron scattering in the low-loss energy regime. Ultramicroscopy 119:51–56. doi: 10.1016/j.ultramic.2011.11.013